\documentclass[a4paper,11pt]{article}

\usepackage{bm}
\usepackage{color}
\usepackage[usenames,dvipsnames,svgnames,table]{xcolor}
\usepackage{amsmath}
\usepackage{amssymb}
\usepackage{graphicx}
\usepackage{slashed}
\usepackage{soul}
\usepackage{multirow}
\usepackage{subfigure}
\usepackage{siunitx} 
\usepackage[utf8]{inputenc}
\usepackage[colorlinks,linkcolor=magenta,citecolor=magenta,urlcolor=magenta]{hyperref}
\usepackage[a4paper,top=2.5cm,bottom=2.5cm,left=2.5cm,right=2.5cm]{geometry}
\usepackage{authblk}
\usepackage{cite}

\newcommand{\rom}[1]{\uppercase\expandafter{\romannumeral #1\relax}}

\let\jnfont=\rm
\def\NPB#1,{{\jnfont Nucl.\ Phys.\ B }{\bf #1},}
\def\PLB#1,{{\jnfont Phys.\ Lett.\ B }{\bf #1},}
\def\EPJC#1,{{\jnfont Eur.\ Phys.\ Jour.\ C }{\bf #1},}
\def\PRD#1,{{\jnfont Phys.\ Rev.\ D }{\bf #1},}
\def\PRL#1,{{\jnfont Phys.\ Rev.\ Lett.\ }{\bf #1},}
\def\MPLA#1,{{\jnfont Mod.\ Phys.\ Lett.\ A }{\bf #1},}
\def\JPG#1,{{\jnfont J.\ Phys.\ G}{\bf #1},}
\def\CTP#1,{{\jnfont Commun.\ Theor.\ Phys.\ }{\bf #1},}
\def\ZPC#1,{{\jnfont Z.\ Phys.\ C }{\bf #1},}
\def\JHEP#1,{{\jnfont JHEP \ }{\bf #1},}
\title{Current Status of Natural NMSSM in Light of LHC 13TeV Data and XENON-1T Results}

\author[a]{Junjie Cao \thanks{junjiec@itp.ac.cn}}
\author[a]{Yangle He \thanks{heyangle@htu.edu.cn}}
\author[a]{Liangliang Shang \thanks{shangliangliang@htu.edu.cn}}
\author[b]{Yang Zhang \thanks{yang.zhang@monash.edu}}
\author[a]{Pengxuan Zhu \thanks{zhupx99@icloud.com}}
\affil[a]{Department of Physics, Henan Normal University, Xinxiang 453007, China}
\affil[b]{ARC Centre of Excellence for Particle Physics at the Tera-scale, School of Physics and Astronomy, Monash University, Melbourne, Victoria 3800, Australia}

\date{}

\begin{document}
    \maketitle

    \begin{abstract}
    In the natural realization of the Next-to-minimal Supersymmetric Standard Model, Higgsinos tend to be lighter than about several hundred GeVs, which can induce detectable leptonic signals at the LHC as well as large DM-nucleon scattering cross section. We explore the constraints from the direct searches for electroweakino and slepton at the LHC~Run~\rom{2} and the latest DM direct detection experiments on the scenario with low fine tuning indicator $\Delta_{Z/h} \leq 50$. We find that  these experiments are complementary to each other in excluding the scenario, and as far as each kind of experiment is concerned, it is strong enough to exclude a large portion of the parameter space. As a result, the scenario with Bino- or Higgsino-dominated DM is disfavored, and that with Singlino-dominated DM is tightly limited. There are two regions in natural NMSSM parameter space surviving in the current experimental limits. One is featured with a decoupled Singlino-dominated LSP with $\mu \simeq m_{\widetilde{\chi}_1^0}$, which cannot be explored by neither DM detections or collider searches. The other parameter space region is featured by  $10^{-47}~{\rm cm^2} \lesssim \sigma^{SI}_{\widetilde{\chi}-p} \lesssim 10^{-46}~{\rm cm^2}$ and the correlation $\mu \simeq m_{\widetilde{\chi}_1^0}$, which will be explored by near future DM detection experiments.
    \end{abstract}
    \newpage
    \section{\label{Introduction}Introduction}
    It is well known that supersymmetric theories provide an elegant solution to the fine tuning problem in the Higgs sector of the Standard Model (SM), where the quadratically divergent contributions to the Higgs mass from the SM fermion loops are canceled exactly by those from corresponding sfermion loops due to supersymmetry, and consequently only relatively mild logarithmic contributions are left in the radiative correction~\cite{Haber:1984rc,Djouadi:2005gj}. This kind of theories also provide the possibility to unify different forces in nature and feasible dark matter (DM) candidates, which must be present in the universe to explain a large number of cosmological and astrophysical observations.  Due to these advantages, supersymmetry has long been regarded as the footstone in building new physics models.

    \par As the most economical realization of supersymmetry in particle physics, the Minimal Supersymmetric Standard Model (MSSM) is theoretically unsatisfactory due to its $\mu$-problem and little hierarchy problem which was firstly discussed in \cite{Barbieri:2000gf,Giudice:2006sn} and became exacerbated in the last few years by the first run of LHC experiments, especially by the uncomfortable large mass of the discovered Higgs boson $m_h \simeq 125~{\rm GeV}$~\cite{Aad:2015zhl}. Alternatively, its gauge singlet extension called the Next-to-Minimal Supersymmetric Standard Model (NMSSM) has drawn a lot of attention since the first hint of the scalar appeared at the LHC~\cite{Ellwanger:2011aa,Gunion:2012zd,Kang:2012sy,King:2012is,Cao:2012fz,Vasquez:2012hn}. In the NMSSM, the $\mu$ parameter is dynamically generated by the vacuum expectation value (vev) of the singlet Higgs superfield $\hat{S}$, and since the field involves in the electroweak symmetry breaking, the magnitude of $\mu$ is naturally at weak scale~\cite{Ellwanger:2009dp,Maniatis:2009re}. Moreover, the interaction among Higgs fields $\lambda \hat{S} \hat{H}_u \cdot \hat{H}_d$ can lead to a positive contribution to the squared mass of the discovered Higgs boson, and if the boson corresponds to the next-to-lightest CP-even Higgs state, its mass can be further enhanced by the singlet-doublet Higgs mixing. These effects make the large radiative correction to the mass unnecessary and thus avoid the little hierarchy problem~\cite{Ellwanger:2011aa, Kang:2012sy,Cao:2012fz,Jeong:2012ma,Badziak:2013bda}.

    \par Compared with the MSSM, the introduction of the singlet field $\hat{S}$ has profound impacts on the phenomenology of the NMSSM, which is reflected in at least two aspects. One is that the scalar component fields of $\hat{S}$ will mix with the doublet Higgs fields to form mass eigenstates. Consequently, the properties of the resulting Higgs bosons may deviate significantly from the MSSM predictions~\cite{Cao:2012fz,Choi:2012he}. In particular, the model predicts singlet-dominated scalars, which can be rather light without conflicting with any experimental constraints~\cite{Cao:2013gba,Ellwanger:2015uaz}, and they may act as the mediators or final states of DM annihilation~\cite{Cao:2013mqa}, and/or as the decay product of heavy sparticles~\cite{Cerdeno:2013qta,Ellwanger:2014hia,Chakraborty:2015xia}. The other is that the involvement of the Singlino, the fermionic component field of $\hat{S}$, in the electroweakino sector usually extends the decay chain of sparticles~\cite{Das:2012rr,Potter:2015wsa,Kim:2015dpa}. This case along with the scenario of sparticle decay into the singlet-dominated scalars lead to complicated signal of sparticles at LHC. In the situation that most of the analyses in sparticle search performed by ATLAS and CMS collaborations which are designed for the MSSM, the constraints on the NMSSM can be much weaker~\cite{Das:2012rr,Cerdeno:2013qta,Ellwanger:2014hia,Chakraborty:2015xia,Potter:2015wsa,Kim:2015dpa}. Besides, due to the presence of the light singlet-dominated scalars and the self interaction of singlet fields $\kappa \hat{S}^3$, the Singlino component in the lightest neutralino $\widetilde{\chi}_1^0$ makes it a more flexible DM candidate to escape the restriction from DM direct and indirect detection experiments in broad parameter space~\cite{Cao:2013mqa} as well as to explain exotic signals observed by DM experiments in certain scenarios~\cite{Cao:2015loa,Cao:2014efa,Guo:2014gra,Bi:2015qva}. All these novel features, therefore, necessitate a detailed study of any relevant parameter point in the NMSSM to see whether it is consistent with experimental data.

    \par In the NMSSM, the $Z$ boson mass is given by~\cite{Baer:2012uy}
	\begin{equation}
	\label{mz}
        \frac{m_{Z}^2}{2} = \frac{m_{H_d}^2+\sum_d-(m_{H_u}^2+\sum_u)\tan^2\beta}{\tan^2\beta-1}-\mu^2,
	\end{equation}
    where $m_{H_d}$ and $m_{H_u}$ are the weak scale soft SUSY breaking masses of the Higgs fields $H_d$ and $H_u$, $\sum_d$ and $\sum_u$ denote their radiative corrections, $\mu$ is the Higgsino mass and $\tan\beta={v_u}/{v_d}$ with $v_u$ and $v_d$ being the vevs of the fields $H_u$ and $H_d$. The equation indicates that, in order to get the observed value of $Z$ boson mass $m_Z$ without resorting to large cancellations, each term on its right hand side should be comparable in magnitude to $m_Z$. The extent of the comparability can be measured by the quantity \cite{Ellwanger:2011mu}
	\begin{equation}\label{eqfint:mz}
	    \Delta_{Z}\equiv \max_{i}\left|\frac{\partial\log{m_Z^2}}{\partial\log{p_i}}\right|,
	\end{equation}
    with $p_i $ denoting any Lagrangian parameter in the NMSSM. Obviously, the smaller value $\Delta_Z$ takes, the more natural the theory is in predicting $m_Z$. On the other hand, any upper bound on $\Delta_Z$ has non-trivial requirements on the parameter space of the NMSSM, e.g. the Higgsino mass is restricted by $\mu\lesssim 300~{\rm GeV}$ and the lighter top squark is bounded by $m_{\widetilde{t}_1} \lesssim 3~{\rm TeV}$ if $\Delta_{Z}<30$~\cite{Baer:2012uy}. In a similar way, one may define another independent quantitative measure of electroweak naturalness from the expression of the SM-like Higgs boson mass~\cite{Farina:2013fsa}
    \begin{equation}\label{mhfint}
	    \Delta_{h}\equiv \max_{i}\left|\frac{\partial\log{m_h^2}}{\partial\log{p_i}}\right|.
	\end{equation}
	In history, the scenario with $\Delta_Z \lesssim {\cal{O}}(10^2)$ is dubbed as Natural SUSY (NS)~\cite{Barbieri:1987fn} or natural NMSSM so far as the explicit model NMSSM is concerned. In recent years with $m_h$ being measured more and more precisely, $\Delta_h$ is also considered in defining the NS~\cite{Baer:2012uy,Baer:2013gva}. As for the natural NMSSM scenario, it should be noted that the novel features mentioned above still hold, which make it differ greatly from the natural MSSM scenario. It should also be noted that the scenario prefers relatively light Higgsinos and scalar top quarks, and this preference can be tested at the LHC.
	
    \par So far the parameter space of the natural NMSSM has been explored relentlessly by considering the constraints from the on-going collider experiments and DM direct and indirect detection experiments~\cite{Potter:2015wsa,Kim:2015dpa,Hall:2011aa,Perelstein:2012qg,Agashe:2012zq,King:2012tr,Gherghetta:2012gb,Kang:2013rj,Cheng:2013fma,Barbieri:2013hxa,Ellwanger:2013rsa,Kim:2013uxa,Binjonaid:2014oga,Kim:2014noa,Cao:2014kya,Dutta:2014hma,Allanach:2015cia,Beuria:2015mta,Enberg:2015qwa,Li:2015dil,Beuria:2016mur,Cao:2016nix,Xiang:2016ndq,Kim:2016rsd,Cao:2016cnv,Cao:2016uwt,Beskidt:2017xsd,Heng:2017jgh,Ellwanger:2016sur,Athron:2017fxj,Ellwanger:2018zxt}. These studies indicate that, although the experiments are very effective in excluding the parameter points of the scenario,  $\Delta_Z$ and $\Delta_{h}$ may still be as low as 2, and the property of the DM candidate is diverse, e.g. it may be either Bino-, Singlino- or Higgsino-dominated~\cite{Cao:2016nix}. This situation, however, may be changed greatly since experimental search for the production of SUSY particles at LHC and DM-nucleon scattering in DM direct detection experiments has made considerable progress in the last years, which was emphasized in recent works~\cite{Datta:2018lup, Pozzo:2018anw}. For example, compared with the LHC~Run~\rom{1} results, the LHC~Run~\rom{2} data have pushed the mass limits on Wino-like $\widetilde{\chi}_1^{\pm}/\widetilde{\chi}_2^{0}$ from $345~{\rm GeV}$ to $650~{\rm GeV}$ in simplified model with $\widetilde{\chi}_1^{0}=0~{\rm GeV}$~\cite{Sirunyan:2018ubx}, and the recent XENON-1T experiment~\cite{Aprile:2018dbl} has improved the sensitivity of the scattering rate by about three times in comparison with the results obtained in 2017 by LUX and PandaX-II experiments~\cite{daSilva:2017swg,Cui:2017nnn}. So in this work we update previous analyses on the natural NMSSM by including the latest experimental data, and we find that the parameter space of the scenario with $\Delta_{Z/h} \leq 50$ shrinks greatly, i.e. some cases become highly disfavored, while some remaining cases will be explored in near future. With the best of our knowledge, these conclusions are not obtained before.

    \par This paper is organized as follows: in Sec.~\ref{sec:model} we introduce briefly the basics of the NMSSM, and present the results of our exhaustive scans over the scenario with $\Delta_{Z/h}\leq50$ by considering various experiment constraints, including the search for sparticles at the LHC~Run I.  Then we show the impact of latest LHC and DM direct detection constraints on different cases in natural NMSSM in Sec.~\ref{sec:constrain}. The status of the scenario is discussed in Sec.~\ref{sec:status}. Finally, we draw our conclusion in Sec.~\ref{sec:conclusion}.

    \section{\label{sec:model}Model and Scan Strategies}
    \subsection{\label{model}Basics of the NMSSM}
    As the simplest extension of the MSSM, the NMSSM contains one extra gauge singlet Higgs field $\hat{S}$ with the superpotential and soft breaking terms given by~\cite{Ellwanger:2009dp}:
    \begin{eqnarray}
        W&=&W_{F}+\lambda \hat{H}_{u}\cdot \hat{H}_{d} \hat{S} + \frac{1}{3}\kappa\hat{S}^3,  \nonumber \\
        V_{\rm{soft}} &=& m_{H_u}^2 |H_u|^2 + m_{H_d}^2 |H_d|^2 + m_S^2 |S|^2 + (\lambda A_{\lambda} S H_u \cdot H_d + \frac{1}{3} \kappa A_{\kappa} S^3 + h.c.) + \cdots, \nonumber
    \end{eqnarray}
    where $W_{F} $ stands the MSSM superpotential without the $\mu$-term, $\hat{H}_u$, $\hat{H}_d$ and $\hat{S}$ are Higgs superfields with $H_u$, $H_d$ and $S$ being their scalar components respectively, the dimensionless coefficients $\lambda$ and $\kappa$ parameterize the coupling strength in Higgs sector, and the dimensional quantities $m_{H_u,H_d,S}^2$ and $A_{\lambda,\kappa}$ are soft breaking parameters.

    \par The Higgs potential of the model consists of the $F$-term and $D$-term of the superfields, as well as the soft breaking terms. After the electroweak symmetry breaking, the fields $H_u$, $H_d$ and $S$ acquire the vevs $v_u$, $v_d$ and $v_s$, and the soft breaking masses $m_{H_u}^2$, $m_{H_d}^2$ and $m_S^2$ can be expressed in terms of $v_u $, $v_d $ and $v_s $ through the minimization conditions of the scalar potential. In practice, the input parameters of the Higgs sector are usually chosen as
    \begin{equation}
        \lambda, \quad \kappa, \quad \tan{\beta} = \frac{v_u}{v_d}, \quad \mu=\lambda v_s, \quad M_A=\frac{2\mu(A_{\lambda}+\kappa v_s)}{\sin{2\beta}}, \quad A_{\kappa},
    \end{equation}
    instead of the soft masses. Moreover, it is more convenient to consider the field combinations $H_1=\cos{\beta} H_u+\varepsilon \sin{\beta} H_d^*$ and $H_2=\sin{\beta} H_u+\varepsilon \cos{\beta} H_d^*$ ($\varepsilon$ is two-dimensional antisymmetric tensor) in discussion, which take the form~\cite{Ellwanger:2009dp}:
    \begin{eqnarray}
        H_1 =   \left ( \begin{array}{c}
                    H^+ \\
                    \frac{S_1 + \mathrm{i} P_1}{\sqrt{2}}
                \end{array} \right),~~
        H_2 &=& \left ( \begin{array}{c}
                    G^+ \\
                    v + \frac{S_2 + \mathrm{i} G^0}{\sqrt{2}}
                \end{array} \right),~~
        H_3  =  v_s +\frac{1}{\sqrt{2}} \left( S_3 + \mathrm{i} P_2 \right),
    \end{eqnarray}
    with $G^+$ and $G^0$ corresponding Goldstone bosons and $v^2 = v_u^2 + v_d^2$. In the basis $(S_1,S_2,S_3)$, the $3\times 3$ symmetric CP-even Higgs mass matrix $M^2$ is given by
    \begin{eqnarray}
        M_{S_1S_1}^2 &=&  M^2_A + (m^2_Z -\lambda^2 v^2) \sin^2 2\beta, \quad M_{S_1S_2}^2 =  -\frac{1}{2}(m^2_Z-\lambda^2 v^2)\sin4\beta, \nonumber \\
        M_{S_1S_3}^2 &=&  -(\frac{M^2_A}{2\mu/\sin2\beta}+\kappa v_s) \lambda v\cos2\beta, \quad M_{S_2S_2}^2 =  m_Z^2\cos^2 2\beta +\lambda^2v^2\sin^2 2\beta, \nonumber \\
        M_{S_2S_3}^2 &=&  2\lambda\mu v[1-(\frac{M_A}{2\mu/\sin2\beta})^2 -\frac{\kappa}{2\lambda}\sin2\beta], \nonumber \\
        M_{S_3S_3}^2 &=&  \frac{1}{4}\lambda^2 v^2(\frac{M_A}{\mu/\sin2\beta})^2 +\kappa v_s A_{\kappa}+4(\kappa v_s)^2 -\frac{1}{2}\lambda\kappa v^2 \sin 2\beta, \nonumber
    \end{eqnarray}
    and consequently the mass eigenstate of CP-even Higgs bosons is $h_{i} = \sum_j V_{ij} S_j$ with $V$ denoting the rotation matrix to diagonalize the mass matrix $M^2$. In a similar way, one can get the CP-odd mass eigenstates $A_1$ and $A_2$. In the following, we take $m_{h_1}<m_{h_2}<m_{h_3} $ and $m_{A_1} < m_{A_2} $, and call $h_i $ the SM-like Higgs boson if its dominant component is the $S_2$ field. Without the mixing of the $S_i$ fields, the squared mass of SM-like Higgs boson gets an additional contribution $\lambda^2 v^2 \sin^2 2\beta$ in comparison with that of MSSM (see the expression of $M_{S_2S_2}^2$), and it can be further enhanced by the mixing effect if $M_{S_3S_3}^2 < M_{S_2S_2}^2$. Consequently, the SM-like Higgs boson does not need large radiative correction to get its measured mass value~\cite{Cao:2012fz,Jeong:2012ma,Badziak:2013bda}. We remind that current experiments have very weak constraints on the $S_3/P_2$ dominated scalars, and as a result, these particles may be as light as several GeVs.

    \par At this stage, it is necessary to clarify that the parameter $p_i$ in Eq.~(\ref{eqfint:mz}) actually denotes the set of the parameters $m_{H_u}^2$, $m_{H_d}^2$, $m_S^2$, $\lambda$, $\kappa$, $A_\lambda$, $A_\kappa$ and $Y_t$ since by the definition of $\Delta_Z$, $m_Z$ should be treated as a variable instead of a constant~\cite{Ellwanger:2011mu}. In this case, $m_Z$, $\tan \beta$ and $\mu$ depend on the Lagrangian parameters by the minimization conditions, which enables one to get their derivatives in an analytic formula ~\cite{Ellwanger:2011mu}. Similar treatment is applied to the calculation of $\Delta_h$ by noting that $m_h$ is related with the parameters by the secular equation $\det \left( M^2 - m_h^2 I_3 \right) = 0 $ ($I_3$ denotes a $3 \times 3$ identity matrix) since $m_h^2$ is one of the eigenvalues of the squared mass matrix $M^2$, and the minimization conditions~\cite{Farina:2013fsa}.

    \par In the NMSSM, the Singlino $\widetilde{S}$ mixes with the gauginos (denoted by $\widetilde{B}$ and $\widetilde{W}$ respectively) and the Higgsinos $\widetilde{H}_d^0$ and $\widetilde{H}_u^0$ to form five neutralinos. In the basis of $\psi = (-i\widetilde{B}, -i\widetilde{W}^3, \widetilde{H}_d^0, \widetilde{H}_u^0, \widetilde{S})$, the symmetric mass matrix $\mathcal{M}_0$ is given by~\cite{Ellwanger:2009dp}
	\begin{equation}
	\mathcal{M}_0=\begin{pmatrix}
 		M_1	&0		&-\frac{g_1 v_d}{\sqrt{2}}	&\frac{g_1v_u}{\sqrt{2}}	&0\\
 			&M_2	&\frac{g_2v_d}{\sqrt{2}}	&-\frac{g_2v_u}{\sqrt{2}}	&0\\
 			& 		&0							&-\mu 						&-\lambda v_u\\
 			& 		& 							&0 							&-\lambda v_d\\
 			& 		& 							& 							&2\kappa v_s
 	\end{pmatrix},
	\end{equation}
    where $M_1$ and $M_2$ are soft breaking masses of Bino and Wino fields respectively, and $g_1$ and $g_2$ are SM gauge couplings. This matrix can be diagonalized by a rotation matrix $N$ so that the mass eigenstates $\widetilde{\chi}_i^0$ are given by
    \begin{eqnarray}\label{eq:netralino_matrix}
        \widetilde{\chi}_i^0 = \sum_{j=1}^5 N_{ij} \psi_j,
    \end{eqnarray}
    where $m_{\widetilde{\chi}_i^0}$ is arranged in ascending order of mass, and thus $\widetilde{\chi}_1^0$ corresponds to DM candidate. The matrix element $N_{ij}$ measures the component of $\psi_j$ field in neutralino $\widetilde{\chi}_i^0$, and we call the DM to be $\psi_j$ dominated if $N_{1j}^2$ is larger than the other components. Note that if any two of the five fields are decoupled, one can get the analytic forms of $N_{1j}$~\cite{Cao:2015loa}, which are useful to understand intuitively DM physics.

    The properties of the other sparticles, such as their masses, are same as those predicted by the MSSM except that they may couple with the singlet fields, which may make their decay product quite complicated and thus increase degree of difficulty in probing them at the LHC~\cite{Das:2012rr,Potter:2015wsa,Kim:2015dpa,Cao:2016nix}. As a result, the exclusion capability of the LHC on the parameter space of the NMSSM is usually weaker than that on the parameter space of the MSSM.
	
    \subsection{\label{sec:scan}Features of Natural NMSSM}

    In order to show in detail the features of the natural NMSSM, we repeat the calculation of our previous works~\cite{Cao:2016nix,Cao:2016cnv} to get more parameter points than what we obtained in these works. Roughly speaking,  we first fix the soft breaking parameters for first two generation squarks and gluino mass at $2~{\rm TeV}$, set a common value $M_{\widetilde{\ell}}$ for all soft breaking parameters in slepton sector, and assume $M_{U_3}=M_{D_3} $ and $A_{t}=A_{b} $ for the third generation squark section to decrease the number of free parameters. Then we scan by Markov Chain method the rest parameters as follows
    \begin{equation}\label{eq:input}
    \begin{split}
        &0<\lambda<0.75,\quad \left| \kappa \right|<0.75,\quad 2<\tan{\beta}<60,\quad 100~{\rm GeV}\leq M_{\widetilde{\ell}}\leq 1.2~{\rm TeV},\\
        &100~{\rm GeV}\leq \mu \leq 1~{\rm TeV},\quad 50~{\rm GeV} \leq M_A \leq 2~{\rm TeV},\quad \left|A_\kappa\right| \leq 2~{\rm TeV}, \\
        &100~{\rm GeV} \leq M_{Q_3}, M_{U_3} \leq 2~{\rm TeV}, \quad \left|A_t\right| \leq \min(3\sqrt{M_{Q_3}^2 + M_{U_3}^2},5~{\rm TeV}),\\
        & \left| M_1\right| \leq 800~{\rm GeV}, \quad 100~{\rm GeV}\leq M_2 \leq 1.2~{\rm TeV},
    \end{split}
    \end{equation}
    with all the parameters defined at the scale of 1 TeV. In the calculation, the particle spectrum is generated by the package \texttt{NMSSMTools}~\cite{Ellwanger:2004xm, Ellwanger:2005dv}, the DM relic density and its spin-independent (SI) and spin-dependent (SD) cross sections are computed with the package \texttt{micrOMEGAs}~\cite{Belanger:2008sj,Belanger:2010gh}, and the likelihood function is taken same as that in \cite{Cao:2018iyk} except that the limits of LUX-2016 for SI cross section~\cite{Akerib:2016lao} and LUX-2016 for SD cross section~\cite{ Akerib:2016vxi}, instead of the limits of the latest XENON-1T results~\cite{Aprile:2018dbl}, are adopted since we are going to show the impact of the latest DM detection experiments on the scenario.  Note that we take the convention $M_2 > 0$ in the scan and allow $M_1$ and $\kappa$ to be either positive or negative. We keep $\mu M_2 > 0$ since this usually leads to a positive contribution from sparticle loops to muon anomalous magnetic moment, which is helpful to alleviate the discrepancy between the measured value of the moment and its SM prediction (we will discuss this issue later)~\cite{Endo:2013bba}.  Due to the differences, the parameter region considered in this work is much broader than that in~\cite{Cao:2016nix,Cao:2016cnv}.

    We further refine the samples obtained in the scan by picking up those which satisfy $\Delta_{Z} \leq 50 $, $\Delta_{h} \leq 50$ and all the constraints implemented in the \texttt{NMSSMTools}, including various B-physics observables in corresponding experimentally allowed range at $2\sigma$ level,  DM relic density within $\pm 10\%$ around its measured central value $\Omega h^2 = 0.1187$~\cite{Aghanim:2018eyx} \footnote{Note that $10\%$ here denotes the theoretical uncertainties in calculating the density, which are much larger than the uncertainty of the Planck measurement.}, and the upper bounds of LUX-2016 on DM-nucleon scattering cross section at $90\%$ confidence level (C.L.). We also consider the constraints from the direct search for Higgs bosons at LEP, Tevatron and LHC with the package \texttt{HiggsBounds}~\cite{Bechtle:2008jh,Bechtle:2011sb} and perform the 125 GeV Higgs data fit with the package \texttt{HiggsSignals}~\cite{Bechtle:2013xfa,Bechtle:2014ewa,Stal:2013hwa}. Moreover, we implement the constraints from various searches for SUSY at LHC~Run~\rom{1} by following procedure: we firstly use the packages \texttt{FastLim}~\cite{Papucci:2014rja} and \texttt{SModelS}~\cite{Ambrogi:2017neo, Kraml:2013mwa} to obtain preliminary constraints, and then use the package \texttt{CheckMATE}~\cite{Kim:2015wza,Dercks:2016npn,Drees:2013wra} with all published analyses to limit the rest samples. The Monte Carlo events of relevant SUSY processes are generated by the package \texttt{MadGraph5\_aMC@NLO}~\cite{Alwall:2014hca,Alwall:2011uj,Frederix:2018nkq} with the package \texttt{PYTHIA}~\cite{Sjostrand:2006za, Sjostrand:2014zea} for parton showering and hadronization.
	\begin{figure}[ht]
	\centering
	\subfigure[Fine tuning indicators] {\includegraphics[width=0.49\textwidth]{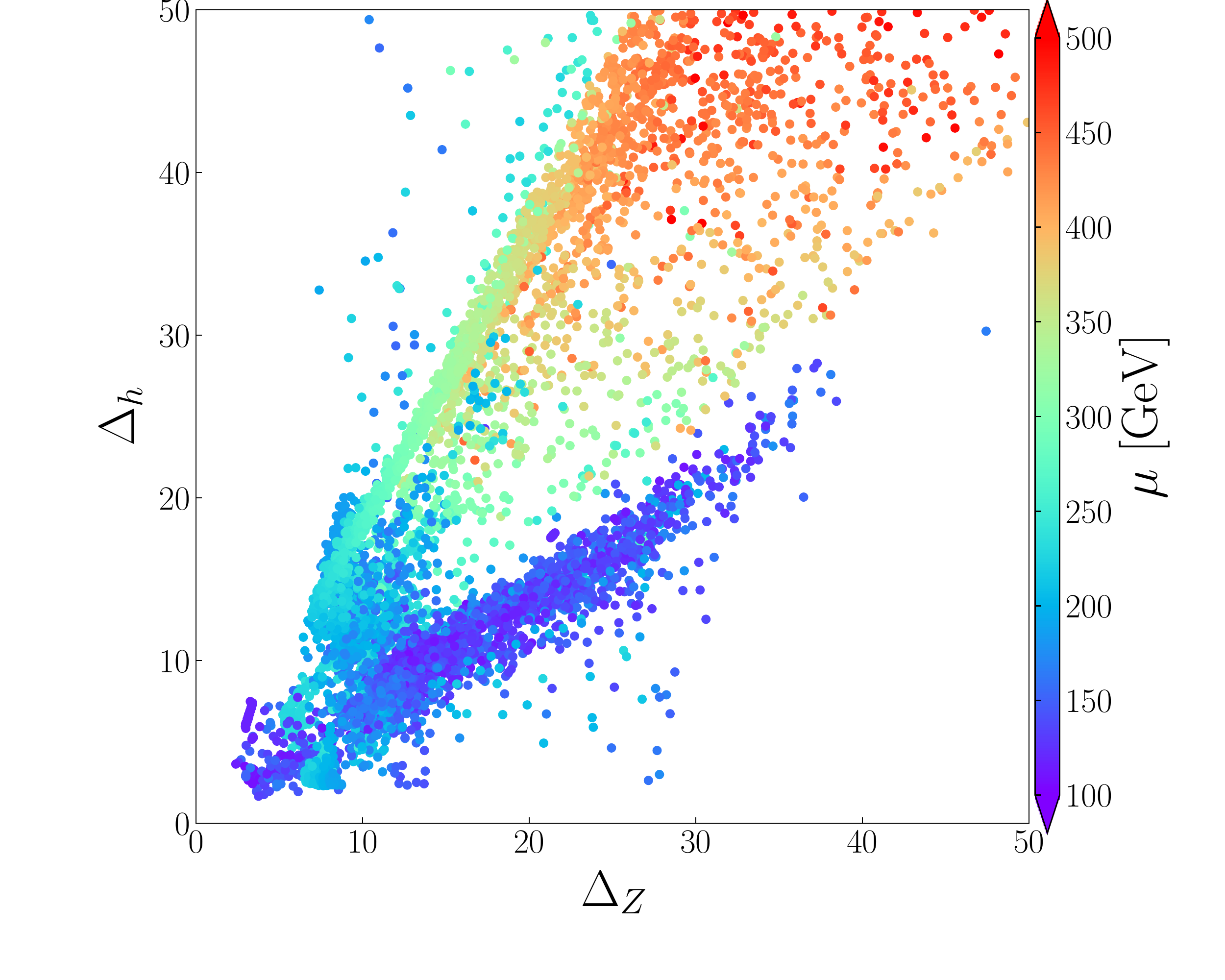}}
	\subfigure[Bino-dominated DM] {\includegraphics[width=0.49\textwidth]{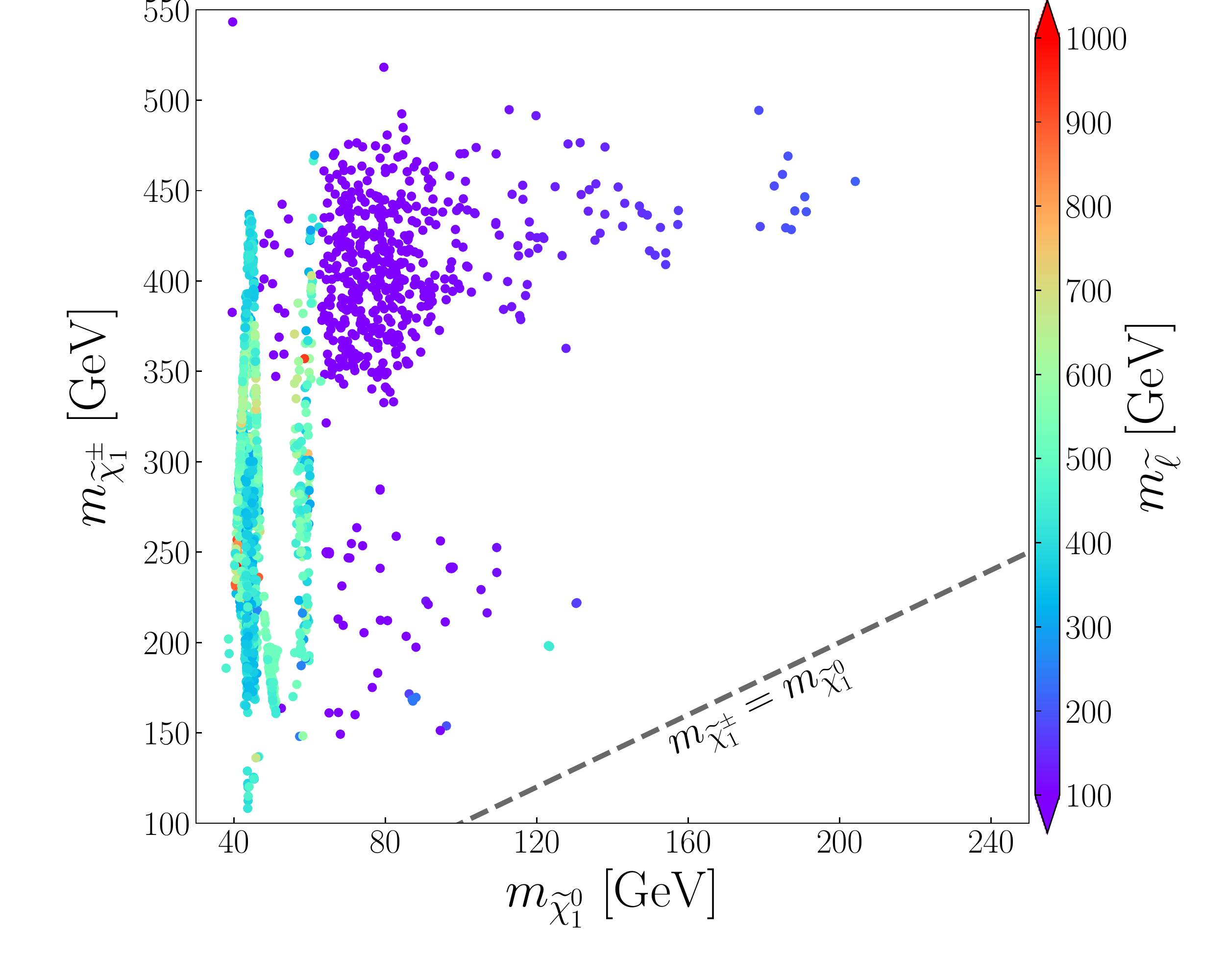}}\\
	\subfigure[Singlino-dominated DM] {\includegraphics[width=0.49\textwidth]{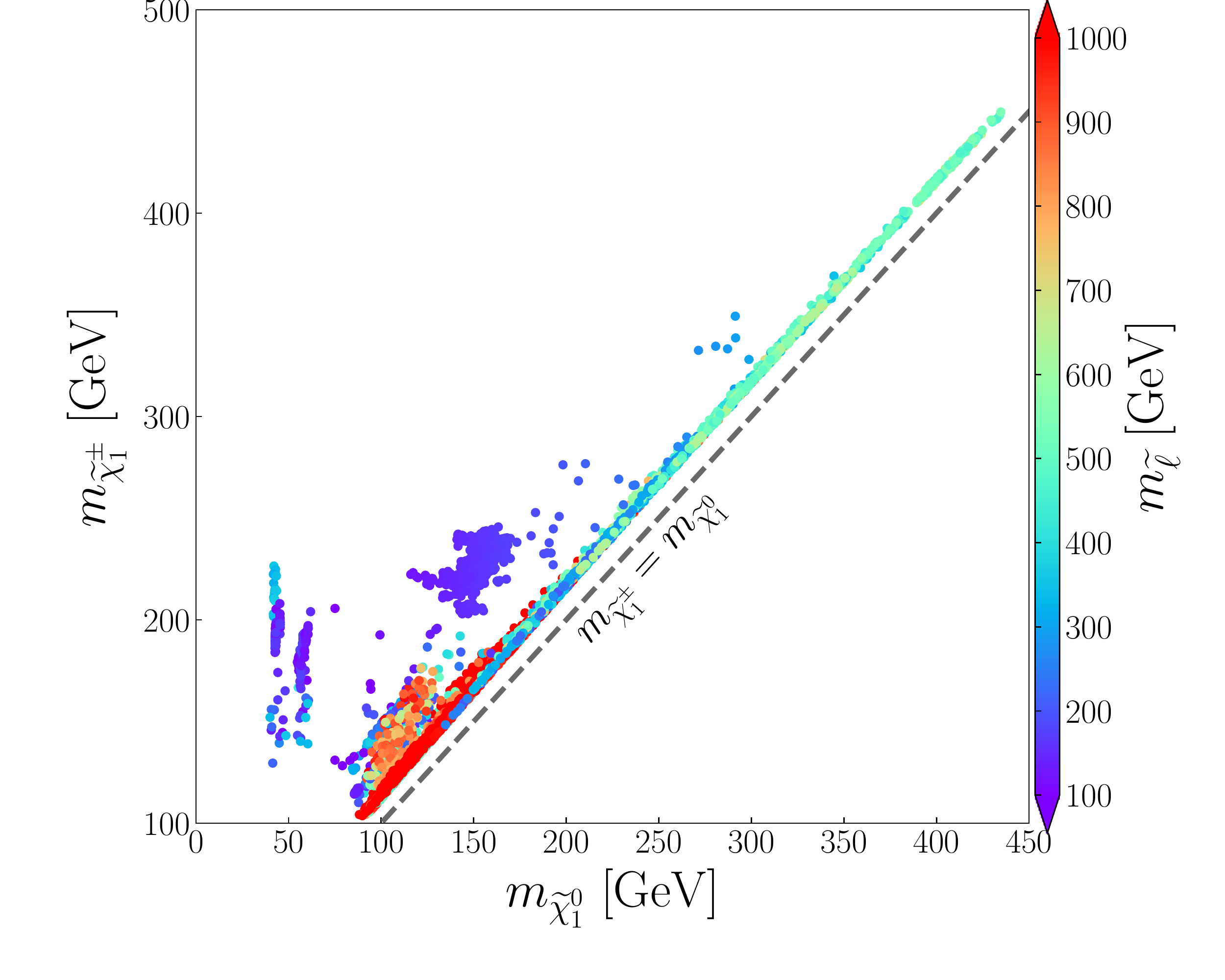}}
	\subfigure[Higgsino-dominated DM] {\includegraphics[width=0.49\textwidth]{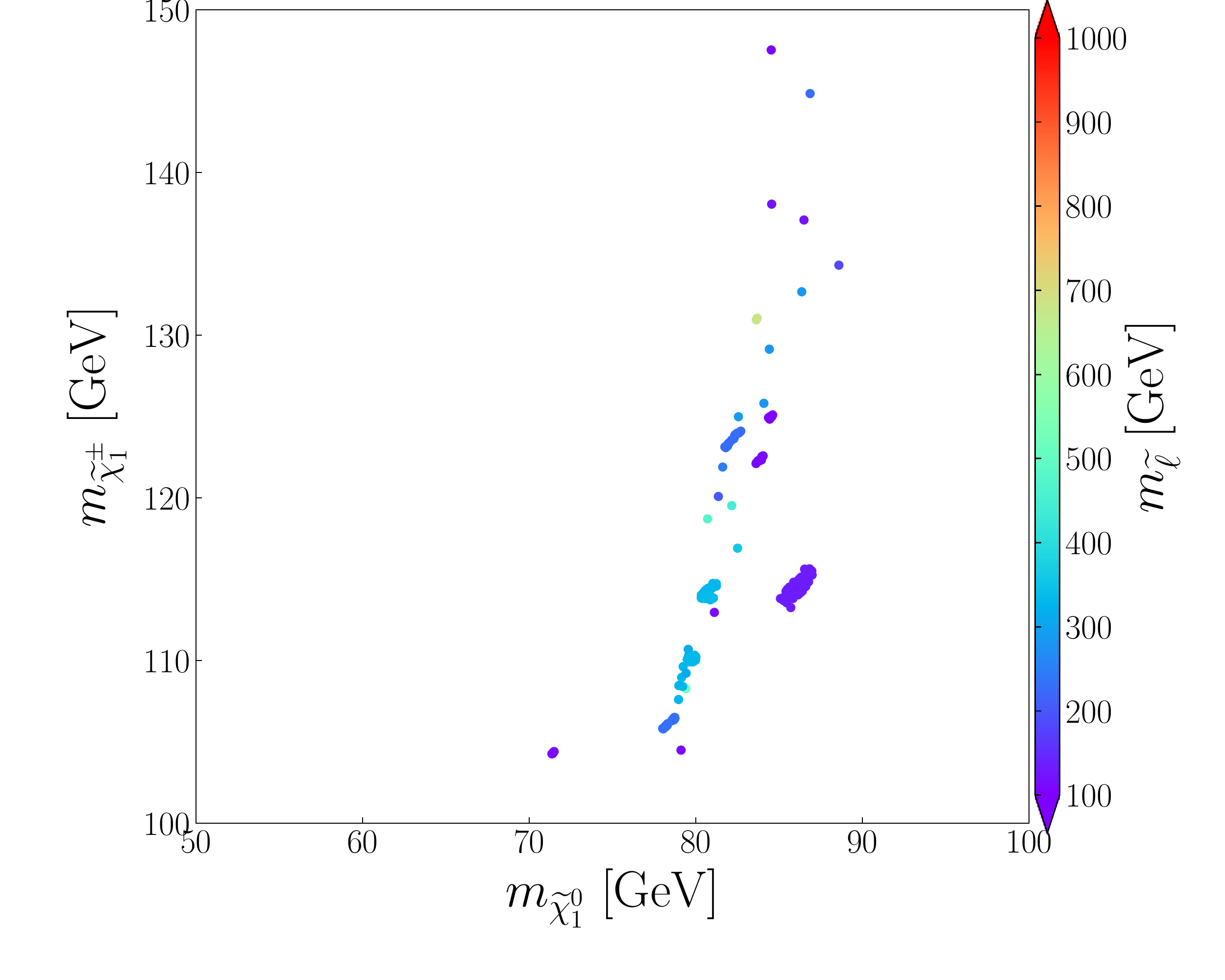}}
	\caption{\label{fig:before} Samples satisfying constraints described in Sec.~\ref{sec:scan} before implementing the latest LHC~Run~\rom{2} and DM direct detection experimental limits. Panel (a) shows the fine tuning indicators $\Delta_{Z} $ versus $\Delta_{h} $. Panel (b), (c) and (d) display the cases of Bino-, Singlino- and Higgsino-dominated DM respectively, on $m_{\widetilde{\chi}_1^0} - m_{\widetilde{\chi}_1^{\pm}}$ plane. The colors indicate the Higgsino mass $\mu$ in panel (a), and the slepton mass $m_{\widetilde{\ell}}$ in panel (b), (c) and (d).}
	\end{figure}

    The scan results before implementing the latest LHC~Run~\rom{2} and DM direct detection experimental limits are presented in Fig.~\ref{fig:before}. In panel (a), we project the samples on the fine tuning indicators $\Delta_Z - \Delta_h$ plane with colors indicating the value of Higgsino mass $\mu$. One can see that $\Delta_{Z} $ and $\Delta_{h} $ can be as low as about 1.7, and $\Delta_{Z/h} \leq 50 $ set an upper limit of 547 GeV on the Higgsino mass $\mu$. This conclusion has been obtained in our previous works~\cite{Cao:2016nix,Cao:2016cnv}, where we aimed to emphasize the importance of the LHC~Run~\rom{1} results and DM direct detection results in limiting the scenario.
    Moreover, in~\cite{Cao:2016cnv} we classified the surviving samples by the dominant component of DM into three types, i.e. Bino-, Singlino- and Higgsino-dominated DM respectively, and found that they show different behaviors to accommodate the constraints from DM detection experiments. In the following, we explore in more detail the features of these types of samples.

    \begin{enumerate}
    \item Bino-dominated DM

    \par For this type of samples, the DM annihilated mainly through three channels to get its measured relic density, which are
    \begin{itemize}
    \item $s$-channel exchange of a resonant SM-like Higgs boson $h_1$ or $Z$ boson\footnote{With above assumptions, namely Bino-like DM and resonant Higgs annihilation, we found only few samples in the scan that predict $h_2$ as the SM-like Higgs boson.}.

    \par In this case, the annihilation cross section is given by~\cite{Ellwanger:2009dp}
    \begin{equation}
    \begin{split}	
    \sigma(\widetilde{\chi}_1^0\widetilde{\chi}_1^0 \overset{h_1}{\to} X X^\prime) &\propto \left|\frac{C_{h_1\widetilde{\chi}_1^0\widetilde{\chi}_1^0} C_{h_1 X X^{\prime}}}{s - m_{h_1}^2 + \mathrm{i} \Gamma_{h_1}m_{h_1}}\right|^2
	    f_s(s, m_{\widetilde{\chi}_1^0}^2, m_X^2, m_{X^{\prime}}^2),\\
	    \sigma(\widetilde{\chi}_1^0 \widetilde{\chi}_1^0 \overset{Z}{\to} X X^{\prime}) &\propto
	    \left|\frac{C_{Z\widetilde{\chi}_1^0\widetilde{\chi}_1^0}C_{Z X X^{\prime}}}{s - m_{Z}^2 + \mathrm{i} \Gamma_{Z} m_{Z}}\right|^2
	    g_s(s, m_{\widetilde{\chi}_1^0}^2, m_X^2, m_{X^{\prime}}^2, m_{Z}^2),  \label{annihilation-funel}
    \end{split}
    \end{equation}
    where $X$ and $X^{\prime}$ denote SM particles, $\Gamma_{h_1}$ ($\Gamma_{Z}$) is the width of $h_1$ ($Z$) boson, $C_{h_1\widetilde{\chi}_1^0\widetilde{\chi}_1^0}$ ($C_{Z\widetilde{\chi}_1^0\widetilde{\chi}_1^0}$) is the coupling between $\widetilde{\chi}_1^0$s and $h_1$ ($Z$) given by
    \begin{equation}
    \begin{split}
        C_{h_1\widetilde{\chi}_1^0\widetilde{\chi}_1^0} & \simeq \sqrt{2}\lambda N_{13}N_{15} - g_1 N_{11}N_{14} + g_2 N_{12} N_{14},\\
        C_{Z\widetilde{\chi}_1^0\widetilde{\chi}_1^0} &= \frac{g_2}{2 \cos{\theta_{W}}}
			\left(-|N_{13}|^2+|N_{14}|^2\right), \label{Couplings}
    \end{split}
    \end{equation}
    and $f_s$ ($g_s$) is the generic functions for $h_1$ ($Z$) funnel depending on the $s$-channel momentum and the involved masses~\cite{Nihei:2002ij}.
    If one further assumes that the Wino and Singlino fields decouple from the rest of the neutralino sector, $N_{12}, N_{15} \sim 0$,
    the other component coefficients of the DM roughly satisfy~\cite{Cao:2015loa}
	\begin{equation}\label{eq:bino}
        N_{11}:N_{13}:N_{14}\simeq
        (m_{\widetilde{\chi}_1^0}^2-\mu^2):
        -\frac{g_1}{\sqrt{2}}(v_u\mu+v_d m_{\widetilde{\chi}_1^0}):
        \frac{g_1}{\sqrt{2}}(v_d\mu+v_u m_{\widetilde{\chi}_1^0}). \nonumber
	\end{equation}
    This relation implies that $|N_{11}| \sim {\cal{O}}(1)$, $N_{13} \propto v_u/\mu$ and $N_{14} \propto (v_d \mu + v_u m_{\widetilde{\chi}_1^0})/\mu^2$ given that $\tan \beta \gg 1$ and $\mu \gg  m_{\widetilde{\chi}_1^0}$, and consequently $C_{h_i\widetilde{\chi}_1^0\widetilde{\chi}_1^0}$ and $C_{Z\widetilde{\chi}_1^0\widetilde{\chi}_1^0}$ are suppressed by $\mu^{-1}$ and $\mu^{-2}$ respectively. Since the annihilation cross section in Eq.~(\ref{annihilation-funel}) must be moderately large to get right DM relic density, $\mu$ should be upper bounded by certain values for the two annihilation channels.

    \par In Fig.~\ref{fig:before}~(b), we show the surviving samples with Bino-dominated DM on $m_{\widetilde{\chi}_1^0} - m_{\widetilde{\chi}_1^{\pm}}$ plane with colors indicating slepton mass $m_{\widetilde{\ell}}$. From this figure, one can see clearly that $\mu \lesssim 480~{\rm GeV}$ and $\mu \lesssim 440~{\rm GeV}$ for the Higgs funnel and $Z$ funnel region respectively (we checked that the lighter chargino is Higgsino dominated for $m_{\widetilde{\chi}_1^\pm} \gtrsim 400~{\rm GeV}$). This situation is similar to the case of MSSM,  which, according to the recent study of~\cite{Pozzo:2018anw}, is strictly limited by the latest LHC~Run~\rom{2} result for electroweakino searches.

    \item $\widetilde{\chi}_1^0\widetilde{\chi}_1^0 \to h_1 h_2$ through $t$-channel exchange of a neutralino $\widetilde{\chi}_i^0$ with $h_2$ corresponding to the SM-like Higgs boson.

    \par This annihilation cross section can be written as
    \begin{equation}
	    \sigma(\widetilde{\chi}_1^0\widetilde{\chi}_1^0 \overset{\widetilde{\chi}_i^0}{\to} X X^\prime) \propto C_{h_1\widetilde{\chi}_i^0\widetilde{\chi}_1^0}^2 C_{h_2 \widetilde{\chi}_i^0\widetilde{\chi}_1^0}^2\
	    h_s(s, m_{\widetilde{\chi}_1^0}^2, m_{\widetilde{\chi}_i^0}^2, m_{h_1}^2, m_{h_2}^2)	
	\end{equation}
    where $C_{h_i \widetilde{\chi}_j^0 \widetilde{\chi}_1^0}$ is the coupling of $h_i$ with $\widetilde{\chi}_1^0 \widetilde{\chi}_j^0$ state, and $h_s$ denotes an auxiliary function encoding the complex mass dependence~\cite{Nihei:2002ij}. We checked that only few samples in Fig.~\ref{fig:before}~(b), which are characterized by $m_{\widetilde{\chi}_1^0}\simeq 100~{\rm GeV}$, $m_{\widetilde{\chi}_1^\pm}\simeq 180~{\rm GeV}$ and $m_{\widetilde{\chi}_1^0}\simeq (m_{h_1}+m_{h_2})/2$, predict the DM to annihilate significantly in this way.

    \item Co-annihilation with sleptons.

    \par With the assumption of a common slepton mass scale $m_{\widetilde{\ell}}$, this annihilation cross section depends only on $m_{\widetilde{\chi}_1^0}$ and $m_{\widetilde{\ell}}$~\cite{Fukushima:2014yia}. As indicated by Fig.~\ref{fig:before}~(b), such co-annihilation channel occurs over a broad range of $m_{\widetilde{\chi}_1^0}$ from $40~{\rm GeV}$ to $220~{\rm GeV}$, and numerical results show the difference of the two masses varying from $60~{\rm GeV}$ to $5~{\rm GeV}$ with the increase of $m_{\widetilde{\chi}_1^0}$. Moreover, we note that either $h_1$ (in most case) or $h_2$ may act as the SM-like Higgs boson in this case.

    \end{itemize}

    \item Singlino-dominated DM
    \par For this type of samples, only $h_2$ ($h_1$) can act as the SM-like Higgs boson for $m_{\widetilde{\chi}_1^0} \lesssim 150 {\rm GeV}$ ($m_{\widetilde{\chi}_1^0} \gtrsim 220 {\rm GeV}$).
    The properties of the DM differ from those of the Bino-like DM in following aspects:

    \begin{itemize}

    \item Besides the three channels for the Bino-dominated DM, the Singlino-dominated DM may also annihilate by the process $\widetilde{\chi}_1^0\widetilde{\chi}_1^0 \to W^+ W^-$ through $t$-channel exchange of a chargino $\widetilde{\chi}_1^\pm$. This requires the mass splitting between $\widetilde{\chi}_1^\pm$ and $\widetilde{\chi}_1^0$ to be about  $10~{\rm GeV} $ for Higgsino-dominated $\widetilde{\chi}_1^\pm$ and about $45~{\rm GeV}$ for Wino-dominated $\widetilde{\chi}_1^\pm$, which is shown on  $m_{\widetilde{\chi}_1^0} - m_{\widetilde{\chi}_1^{\pm}}$ plane in Fig.~\ref{fig:before}~(c) for Singlino-dominated DM case. We note that for the Higgsino case, the co-annihilation of the Higgsinos with $\widetilde{\chi}_1^0$ is also important since the mass splitting is less than $10\%$~\cite{Griest:1990kh,Baker:2015qna}.

    \item For the Singlino-dominated DM, the elements of the matrix $N$ have following relationship~\cite{Cao:2015loa,Cao:2016nix}:
	\begin{equation}
        N_{13}:N_{14}:N_{15}\simeq \lambda(v_d\mu-v_u m_{\widetilde{\chi}_1^0}):\lambda(v_u\mu-v_d m_{\widetilde{\chi}_1^0}):(m_{\widetilde{\chi}_1^0}^2-\mu^2).
	\end{equation}
    in the limit of $|\mu|\ll |M_1|,|M_2|$. This implies that the DM has less Higgsino components than the Bino-dominated DM, and consequently the $h \widetilde{\chi}_1^0\widetilde{\chi}_1^0$ and $Z\widetilde{\chi}_1^0\widetilde{\chi}_1^0$ coupling strengths may be significantly smaller than those for the Bino-dominated DM case if $m_{\widetilde{\chi}_1^0}$, $\mu$, $\lambda$ and $\tan \beta$ are taken same values (see the expressions in Eq.~(\ref{Couplings})). That is why the Higgsino masses in Fig.~\ref{fig:before}~(c) are visibly smaller than that in Fig.~\ref{fig:before}~(b) for the funnel regions.

    \item Compared with the Bino-dominated DM case, we found more samples that the DM annihilate significantly by the channel $\widetilde{\chi}_1^0\widetilde{\chi}_1^0 \to h_1 h_2$. The underlying reason is that the Singlino-dominated DM prefers certain parameter space of the NMSSM, such as $2 |\kappa| < \lambda$ and moderately light $\mu$, so that $h_2$ prefers to be the SM-like Higgs boson for $m_{\widetilde{\chi}_1^0} \sim 100~{\rm GeV}$. By contrast, for Bino-like DM case $h_1$ prefers to be the SM-like Higgs boson.
    \item The DM may annihilate by a resonant singlet-dominated $A_1$, which has long been considered as a viable annihilation mechanism in literature~\cite{Cao:2011re}, but after considering the constraints such a case becomes rare. This fact can be understood from the sum rule for the masses of the singlet fields~\cite{Ellwanger:2016sur,Ellwanger:2018zxt}
    \begin{equation}
        \mathcal{M}_{0,55}^2 \simeq M_{S_3 S_3}^2 + \frac{1}{3} M_{P_2 P_2}^2,
    \end{equation}
    and the approximation $M_{S_3 S_3}^2 \simeq m_{h_1}^2$ and $M_{P_2 P_2}^2 \simeq m_{A_1}^2$, which is valid for most cases. Then the resonant annihilation condition $m_{A_1} \simeq 2 m_{\widetilde{\chi}_1^0}$  implies that
    \begin{equation}
        m_{h_1}^2 \simeq  \mathcal{M}_{0,55}^2 - \frac{4}{3} m_{\widetilde{\chi}_1^0}^2.
    \end{equation}
    Since $m_{h_1}^2 > 0$, the equation holds only when $ \mathcal{M}_{0,55}^2$ is significantly larger than $m_{\widetilde{\chi}_1^0}$, which can be achieved by a large $\lambda$ to induce sizable mixing between Higgsinos and Singlino in the neutralino mass matrix. Such a parameter space then predicts  a light $h_1$ as well as a singlino-dominated DM whose Higgsino component is also sizable. Obviously, this situation tends to predict a large DM-nucleon scattering rate, and is therefore limited by DM direct detection experiments~\cite{Ellwanger:2016sur,Mou:2017sjf,Shang:2018dja,Ellwanger:2018zxt}. In fact, we find that only when the DM mass lies within a range from $88~\rm GeV$ to $122~\rm GeV$ can the situation survive the constraint.

    \par Moreover, we note that some samples with $m_{\widetilde{\chi}_1^0}<10~{\rm GeV}$ and the $A_1$ funnel as DM dominant annihilation channel are presented in~\cite{Mou:2017sjf}. We checked the properties of these samples and found that they have been excluded by the $3\ell+E_{\rm T}^{\rm miss}$ search at the LHC~\cite{Sirunyan:2017lae}.

    \end{itemize}

    \item Higgsino-dominated DM

	\par This scenarios is characterized by approximately degenerated Higgsinos and Singlino in mass, and consequently $\widetilde{H}_u^0$, $\widetilde{S}$ and $\widetilde{H}_d^0$ components of the DM are comparable in magnitude with the largest one coming from the $\widetilde{H}_u^0$~\cite{Cao:2016nix}. The non-negligible Singlino component $N_{15}^2$, which is around 30\%, can dilute the interactions of the DM with other fields so that DM density can coincide with the measured value of WMAP and Planck experiments~\cite{Cao:2016nix}. We checked that the main annihilation channels of the DM in early universe include $\widetilde{\chi}_1^0\widetilde{\chi}_1^0 \to W^+W^-,~ZZ,~Zh_1,~h_1 h_1,~h_1h_2$, where $h_2$ always denotes the SM-like Higgs for this type of samples, as well as the co-annihilation with sleptons. As was shown in~\cite{Cao:2016cnv}, this scenario is tightly restricted by LUX-2016 on SD cross section for DM-nucleon scattering, and only samples with $m_{\widetilde{\chi}_1^0}\simeq 80~{\rm GeV}$ and $\tan \beta \sim {\cal{O}}(1)$ are experimentally allowed. Part of these features are illustrated in Fig.~\ref{fig:before}~(d) of this work as well as in Fig.~2-4 of \cite{Cao:2016cnv}.

    \end{enumerate}

    \par In summary, the natural NMSSM lives quite well before LHC~Run~\rom{2} and DM direct detection experiments. It has various kinds of DM candidates and abundant annihilation mechanisms, but on the other hand it is usually accompanied by some light sparticles which makes it to be tested at the LHC. In what follows, we study the impact of the latest LHC~Run~\rom{2} and DM direct detection experiments on this scenarios.

    \section{Impact of LHC~Run~\rom{2} and DM direct detection results}
    \label{sec:constrain}

    Due to the requirement on naturalness, the DM is bounded by $m_{\widetilde{\chi}_1^0}< 440~{\rm GeV}$ with either moderately light chargino or light slepton. This feature motivates us to study the direct searches for sleptons and neutralinos/charginos pair production at LHC~Run~\rom{2} and DM direct detections.

    \subsection{\label{lhcsearch}Sparticle searches at LHC~Run~\rom{2}}\label{sec:lhc}

    To implement the LHC~Run~\rom{2} limits on the slepton and electroweakino of samples, we add the following LHC~Run~\rom{2} experimental analyses to \texttt{CheckMATE}:
	\begin{itemize}
	\item The CMS search for electroweakinos in final state with either two or more leptons of the same charge, or with three or more leptons~\cite{Sirunyan:2017lae}. In simple terms, the target processes of this analysis are $pp \to \widetilde{\chi}_i^{\pm} \widetilde{\chi}_j^0$ with different decay models into $2/3/4\ell + E_{\rm{T}}^{\rm{miss}}$ final state. The decay models can be classified into light slepton scenario and heavy slepton scenario. In light slepton scenario, the dominated decay chain of neutralino is $\widetilde{\chi}_i^0 \to \ell \widetilde{\ell} \to \ell^{+}\ell^{-}\widetilde{\chi}_1^0$ with $i > 1$, and main decay chain of chargino is $\widetilde{\chi}_i^{\pm} \to \nu_{\ell}\widetilde{\ell}^{\pm}/\widetilde{\nu_{\ell}}\ell^{\pm} \to \ell^{\pm} \nu_{\ell} \widetilde{\chi}_1^0$. The mass of slepton $m_{\widetilde{\ell}}$ and the flavor of the slepton in the decay chain both directly affect the property of final state. In the heavy slepton scenario, decay models $\widetilde{\chi}_i^0\widetilde{\chi}_j^{\pm} \to (Z\widetilde{\chi}_1^0)(W^{\pm}\widetilde{\chi}_1^0)$ and $\widetilde{\chi}_i^0\widetilde{\chi}_j^{\pm} \to (h\widetilde{\chi}_1^0)(W^{\pm}\widetilde{\chi}_1^0)$ with $Z\to \ell\ell$, $W^{\pm} \to jj/ \ell^{\pm}\nu_{\ell}$ and $h\to \ell\ell$ will lead to two/three-lepton final states. Here $h$ refers to the 125 GeV SM Higgs boson. Our natural NMSSM samples cover both scenarios.

    \begin{table}[t]
        \centering
        \resizebox{\textwidth}{!}{
        \begin{tabular}{c|c|c|c|cccccc}
        \hline
        \multirow{2}{*}{SR} & \multirow{2}{*}{$n_{\rm{bin}}$} & \multicolumn{2}{c|}{\multirow{2}{*}{final state}}                                    & \multicolumn{6}{c}{signal region bins defined by}                    \\ \cline{5-10}
                             &                       & \multicolumn{2}{c|}{}                                                                & $E_{\rm{T}}^{\rm{miss}}$          & $M_{\ell\ell}$ & $M_{\rm{T}}$  & $M_{\rm{T2}}$ & $p^{\rm{T}}_{\ell\ell}$ & $n_{\rm{jet}}$  \\ \hline
        \texttt{SS}                   & 30                    & \multicolumn{2}{c|}{2 same sign leptons}                                                       & $> 60$ & -   & yes & -   & yes  & 0 or 1 \\ \hline
        \texttt{SRA}                  & 44                    & \multirow{2}{*}{3 light leptons}             & with an OSSF pair                        & $>  50$ & yes & yes & -   & -    & -      \\ \cline{1-2} \cline{4-10}
        \texttt{SRB}                  & 6                     &                                              & without OSSF pair                     & $>  50$ & yes & yes & -   & -    & -      \\ \hline
        \texttt{SRC}                  & 18                    & \multirow{3}{*}{2 light leptons with 1 $\tau_{\rm{h}}$ }        & with an OSSF pair                        & $>  50$ & yes & -   & yes & -    & -      \\ \cline{1-2} \cline{4-10}
        \texttt{SRD}                  & 16                    &                                              & with OS pair                          & $>  50$ & yes & -   & yes & -    & -      \\ \cline{1-2} \cline{4-10}
        \texttt{SRE}                  & 12                    &                                              & with SS pair                          & $>  50$ & yes & -   & yes & -    & -      \\ \hline
        \texttt{SRF}                  & 12                    & \multicolumn{2}{c|}{1 light lepton with 2 $\tau_{\rm{h}}$}                                            & $> 50$ & yes & -   & yes & -    & -      \\ \hline
        \texttt{SRG}                  & 5                     & \multirow{5}{*}{4 or more than four leptons} & with $n_{\rm{OSSF}} \geq$ 2 no $\tau_{\rm{h}}$      & yes             & -   & -   & -   & -    & -      \\ \cline{1-2} \cline{4-10}
        \texttt{SRH}                  & 4                     &                                              & with $n_{\rm{OSSF}} < $ 2 no $\tau_{\rm{h}}$         & yes             & -   & -   & -   & -    & -      \\ \cline{1-2} \cline{4-10}
        \texttt{SRI}                  & 4                     &                                              & with 1 $\tau_{\rm{h}}$                            & yes             & -   & -   & -   & -    & -      \\ \cline{1-2} \cline{4-10}
        \texttt{SRJ}                  & 4                     &                                              & with $n_{\rm{OSSF}} \geq 2$ and 2 $\tau_{\rm{h}}$ & yes             & -   & -   & -   & -    & -      \\ \cline{1-2} \cline{4-10}
        \texttt{SRK}                  & 3                     &                                              & with $n_{\rm{OSSF}} <$ 2 and 2 $\tau_{\rm{h}}$     & yes             & -   & -   & -   & -    & -      \\ \hline
        \end{tabular}}
        \caption{ The summarisation of signal region categories defined in the CMS search for electroweakinos in final state with two light leptons of the same charge or with three or more leptons~\cite{Sirunyan:2017lae}. ``OSSF'' ``OS'' and ``SS'' stand ``opposite sigh same flavor'', ``opposite sign'' and ``same sign'' leptons, respectively. $\tau_{\rm h}$ denotes tau-tagged jet. ``yes'' means the corresponding variable is used to category bins. All quantities with mass dimension are given in units of GeV.}\label{tab039}
    \end{table}

	After passing the basic selections, the signal events are categorized into 158 bins which are summarized into 12 signal regions (SRs) categories shown in Tab.~\ref{tab039}.
	The first SR category, \texttt{SS}, is designed to the compressed scenarios in which one of the leptons from the decay chain of neutralino can be very soft, and therefore requires 2 same-sign (SS) leptons. The SR categories requiring three reconstructed leptons can be further classified by the number of $\tau_{\rm{h}}$ candidate. For three-leptons final state without $\tau_{\rm{h}}$, signal events with (without) an opposite-sign same flavor (OSSF) lepton pair are categorized into SR category \texttt{SRA} (\texttt{SRB}).	For three-leptons final state with one $\tau_{\rm{h}}$ candidate, SRs are defined as \texttt{SRC}, \texttt{SRD} and \texttt{SRE} by the signal events with OSSF lepton pair, opposite-sign (OS) lepton pair and (SS) lepton pair respectively. The \texttt{SRF} requires two $\tau_{\rm{h}}$ candidates of three reconstructed leptons. The events with final state of four or more than four leptons are classified into \texttt{SRG} to \texttt{SRK} by the number of OSSF pair $n_{\rm{OSSF}}$ and the number of $\tau_{\rm{h}}$. They aim for the production of a $Z$ boson or $h$ Higgs boson in the decay chain, which finally decays into two light flavor leptons or two taus.

	\item The CMS searches for electroweakinos with compressed mass spectra using events including two soft OS leptons and missing transverse energy~\cite{CMS-PAS-SUS-16-048}. The analysis is conceived to provide sensitivity to the process $pp \to \widetilde{\chi}_2^0\widetilde{\chi}_1^{\pm} \to (\widetilde{\chi}_1^0 W^{*})(\widetilde{\chi}_1^0 Z^{*})$ for mass differences between $\widetilde{\chi}_2^0$ and $\widetilde{\chi}_1^0$ ($\Delta m$) of less than 20 GeV, where $Z^{*}$ and $W^{*}$ stand virtual $Z$ and $W$ bosons. The analysis requires an OS pair of light leptons, moderate $E_{\rm{T}}^{\rm{miss}}$ and at least one jet. No significant excess was reported in the 12 SRs defined based on dilepton invariant mass and $E_{\rm{T}}^{\rm{miss}}$. In simplified model, Wino-like $\widetilde{\chi}_2^0/\widetilde{\chi}_1^{\pm}$ masses masses up to 230 GeV are excluded for $\Delta m$ of 20 GeV. This analysis should be sensitive to the Singlino-dominated DM annihilating through $t$-channel chargino.
	
	\item The CMS search for electroweakinos in events with a lepton, two $b$-tagged jets and  significant imbalance in the transverse momentum~\cite{Sirunyan:2017zss}. This search targets the neutralino and chargino pair production $pp \to \widetilde{\chi}_2^0 \widetilde{\chi}_1^{\pm} \to (\widetilde{\chi}_1^0 h)(\widetilde{\chi}_1^0 W^{\pm})$ with decay models $h\to b\bar{b}$ and $W \to \ell \nu_\ell$. The kinematic variables used in this analysis including $E_{\rm{T}}^{\rm{miss}}$, the invariant mass of the two $b$ jets $M_{b\bar{b}}$, the transverse mass of the lepton-$E_{\rm{T}}^{\rm{miss}}$ system $M_{\rm{T}}$ and the contransverse mass variable
	\begin{equation}
		M_{\rm{CT}} = \sqrt{2 p_{\rm{T}}^{b1} p_{\rm{T}}^{b2} [1+\cos(\Delta\phi_{bb})]},
	\end{equation}
	where $p_{\rm{T}}^{b1}$ and $p_{\rm{T}}^{b2}$ are the transverse momenta of the tow $b$ jets, and $\Delta\phi_{bb}$ is the azimuthal angle between the $b$ jets pair. After requiring $ 90~{\rm GeV} < M_{b\bar{b}} < 150~{\rm GeV}$, $M_{\rm{T}} > 150 ~{\rm GeV}$ and $M_{\rm{CT}}> 170~{\rm GeV}$, two exclusive SRs of $125~{\rm GeV}<E_{\rm{T}}^{\rm{miss}}<200 ~{\rm GeV}$ and $E_{\rm{T}}^{\rm{miss}}>200 ~{\rm GeV}$ are defined to  enhance sensitivity to signal models with different mass spectra. The results show no significant excess in the two SRs, and exclude Wino-like $\widetilde{\chi}_2^0/\widetilde{\chi}_1^\pm$ between 220 and 490 GeV when the $\widetilde{\chi}_1^0$ is massless in simplified model. This analysis should be sensitive to the Bino-dominated DM scenario in which Higgsino-like $\widetilde{\chi}_{2,3}^0$ can decay to $\widetilde{\chi}_{1}^0h$ with large branch ratios.
	
	\item The CMS search for electroweakinos in final states with two leptons consistent with a $Z$ boson and $E_{\rm{T}}^{\rm{miss}}$~\cite{Sirunyan:2017qaj}. This search is designed for both strong and electroweak SUSY production leading to the on-$Z$ signature, by selecting events with exactly one OSSF lepton pair consistent with the $Z$ boson mass, two non $b$-tagged jets consistent with the $W$ boson mass and large $E_{\rm{T}}^{\rm{miss}}$. Two Electroweak-production on-$Z$ SRs, \texttt{HZ} and \texttt{VZ}, were defined with the invariant mass of two jets $M_{jj}$, the variable $M_{\rm{T2}}(\ell\ell)$~\cite{Lester:1999tx,Barr:2003rg} using the two selected leptons and $M_{\rm{T2}}(\ell b\ell b)$ using two combinations of one lepton and one $b$-tagged jet as the visible object. The SRs are then divided into bins in $E_{\rm{T}}^{\rm{miss}}$. The analysis excludes Wino-like $\widetilde{\chi}_2^0/\widetilde{\chi}_1^{\pm}$ masses between approximately 160 and 610 GeV for massless $\widetilde{\chi}_1^0$ with decay branch ratios ${\rm Br}(\widetilde{\chi}_1^{\pm} \to W^{\pm} \widetilde{\chi}_1^0) = {\rm Br}(\widetilde{\chi}_2^0 \to Z \widetilde{\chi}_1^0) = 100\%$. Thus it is sensitive to both the Bino-dominated DM scenario and the Singlino-dominated DM scenario.

	\item The CMS search for sleptons in final states with one OSSF lepton pair, no jet and large missing transverse momentum~\cite{CMS-PAS-SUS-17-009}. This search is optimized on the production of selectron pair and smuon pair in simplified model that ${\rm Br}(\widetilde{\ell}\to \ell \widetilde{\chi}_1^0)=100\%$. In order to suppress $t\bar{t}$ and $WW$ backgrounds, the SR selects events with $20 ~{\rm GeV} < M_{\ell \ell} < 76 ~{\rm GeV}$ or $M_{\ell \ell}>126 ~{\rm GeV}$, $M_{\rm{T2}}(\ell\ell)> 90 ~{\rm GeV}$, no jet with $p_{\rm{T}} > 25 ~{\rm GeV}$ and $E_{\rm{T}}^{\rm{miss}}> 100 ~{\rm GeV}$, and then are divided into 4 bins in $E_{\rm{T}}^{\rm{miss}}$. This analysis probes $\widetilde{e}_{L/R}$ and $\widetilde{\mu}_{L/R}$ masses lower than approximately 450 GeV with $m_{\widetilde{\chi}_1^0}=0 \text{ GeV}$. It should be sensitive the $h/Z$ funnel region in the Bino-dominated DM scenario and the Singlino-dominated DM scenario.
	
	\item The CMS search for stau in the semi-leptonic and all-leptonic final state~\cite{CMS-PAS-SUS-17-002}. This search is targeting for direct $\widetilde{\tau}$ pair production process in final state with two different flavor leptons formed one OS pair, which could be divided into $e\mu$, $e\tau$ and $\mu\tau$ channels. The kinematic variable used in this search to bin SRs include $E_{\rm{T}}^{\rm{miss}}$, $M_{\rm{T2}}$ and $D\zeta$, where $D\zeta$ is defined as:
	\begin{equation}
	D\zeta = P_{\zeta, \rm{miss}} - 0.85 P_{\zeta, \rm{vis}},\quad
	P_{\zeta, \rm{miss}}= {\vec{p}}_{\rm{T}}^{\rm{miss}}\cdot \vec{\zeta}, \quad
	P_{\zeta, \rm{vis}} = ({\vec{p}}_{\rm{T}}(\ell_1) + {\vec{p}}_{\rm{T}}(\ell_2))\cdot \vec{\zeta},
	\end{equation}
	here $\vec{\zeta}$ is the bisector between the direction of the two leptons, ${\vec{p}}_{\rm T}(\ell_1)$ and ${\vec{p}}_{\rm T}(\ell_2)$ are the transverse momenta of two leptons. In this search, signal events are binned into 144 SRs. Since the data from collider are consistent the SM expectations, no mass point in direct $\widetilde{\tau}$ production can be excluded. For a $\widetilde{\tau}$ mass of 90 GeV and a $\widetilde{\chi}_1^0$ of 1 GeV with decay mode ${\rm Br}(\widetilde{\tau} \to \tau \widetilde{\chi}_1^0) = 100\%$, the $95\%$ C.L. upper limit for direct $\widetilde{\tau}$ pair production cross section is up to \SI{0.66}{\pico\barn}.
	
	\item The CMS search for stau pair production in the all-hadronic final state~\cite{CMS-PAS-SUS-17-003}. This search examines events with two hadronically decaying $\tau$ leptons and large $E_{\rm{T}}^{\rm{miss}}$. In this search, the angle between two $\tau_{\rm{h}}$ candidates $\Delta\phi(\tau_1, \tau_2)$, $M_{\rm{T2}}(\tau_1, \tau_2)$, $E_{\rm{T}}^{\rm{miss}}$ and $\Sigma M_{\rm{T}}$ are used in the signal selection criteria to reduce the SM background, where $\Sigma M_{\rm{T}} = M_{\rm{T}}(\tau_1, \vec{p}_{\rm{T}}^{\rm{miss}}) + M_{\rm{T}}(\tau_2, \vec{p}_{\rm{T}}^{\rm{miss}})$.  Three exclusive SRs are used to improve the sensitivity towards signal models with different stau masses. This analysis is most sensitive to a scenario with left-handed stau of around 125 GeV and a massless $\widetilde{\chi}_1^0$.
	\end{itemize}
	
	\par We have submitted the implementations of above analyses to the \texttt{CheckMATE} database. The validations of cut-follows can be found in the website and Appendix~\ref{sec:appendix}, which shows that our simulations agree with the corresponding experimental results within a $20\%$ uncertainty.
	
	\par For the surviving samples described in Sec.~\ref{sec:scan}, we generate MC events of following processes
	\begin{equation*}
	\begin{split}
	    pp\to \widetilde{\chi}_i^{\pm}\widetilde{\chi}_j^{0},&\quad i=1,2;\quad j=2,3,4;\\
        pp\to \widetilde{\chi}_i^{\pm}\widetilde{\chi}_j^{\mp},&\quad i=1,2;\quad j=1,2;\\
        pp\to \widetilde{\chi}_i^{0}\widetilde{\chi}_j^{0},&\quad i=2,3,4;\quad j=2,3,4;\\
        pp\to \widetilde{\ell}_i^{\pm}\widetilde{\ell}_i^{\mp}/\widetilde{\ell}_i^{\pm}\widetilde{\nu}_i/\widetilde{\nu}_i\widetilde{\nu}_i,&\quad i=e,\mu,\tau;
	\end{split}
    \end{equation*}
	at 13 TeV LHC, using \texttt{MadGraph5\_aMC@NLO}~\cite{Alwall:2014hca,Alwall:2011uj,Frederix:2018nkq} with the package \texttt{PYTHIA}~\cite{Sjostrand:2006za, Sjostrand:2014zea} for parton showering and hadronization. Although the cross section of slepton pair production is much smaller than the cross section of electroweakino pair production, two high $p_{\rm{T}}$ leptons from sleptons decay are always appeared in the final state. Process $pp \to \widetilde{\nu}\widetilde{\nu}$, for example, can provide a SS lepton pair with a large $E_{\rm{T}}^{\rm{miss}}$ if sneutrino pair decay through $\widetilde{\nu}\widetilde{\nu} \to (\widetilde{\chi}_1^{\pm} \ell^{\mp})(\widetilde{\chi}_1^{\pm} \ell^{\mp})$, which sensitive to the \texttt{SS} category in analysis~\cite{Sirunyan:2017lae}. And then the events are passed into \texttt{CheckMATE} which includes \texttt{Delphes-3.2.0}~\cite{deFavereau:2013fsa} for detector fast simulation. The cross section are normalized to NLO using \texttt{PROSPINO2}~\cite{Beenakker:1996ed}.
	
    \par We firstly use the \texttt{R} values obtained from \texttt{CheckMATE} to apply the constraints from above searches. Here $\texttt{R} \equiv \max\{(S_{i}-1.96\Delta S_{i})/S_{i,\rm{obs}}^{95}\}$ for individual analysis, where $S_{i}$ is the number of simulated signal events in $i_{\rm{th}}$ SR or bin of the analysis, $\Delta S_{i}$ stands the uncertainty of $S_{i}$ and $S_{i,\rm{obs}}^{95}$ represents the $95\%$ C.L. upper limit of the event number in the SR. The samples that the \texttt{R} value of any above analysis is larger than 1 are deemed to be excluded by searches at LHC~Run~\rom{2} at $95\%$ C.L. in the following text. Then we combine the first four CMS electroweakino searches~\cite{Sirunyan:2017lae,CMS-PAS-SUS-16-048,Sirunyan:2017zss,Sirunyan:2017qaj} though $CLs$ method~\cite{Read:2002hq} with \texttt{RooStats}~\cite{Schott:2012zb}, because the SRs of them are mutually exclusive~\cite{Sirunyan:2018ubx}. We use the likelihood function described in~\cite{Pozzo:2018anw} for the combination, in which relative uncertainties of signal event is assumed to equal $5\%$ and covariance matrices are not included.

    \subsection{DM direct detection}\label{sec:dm}

    Complementary to the LHC experiments, DM direct detection experiments can also limit tightly the natural NMSSM scenario by measuring the SI and SD cross section for DM-nucleon scattering. In the NMSSM with heavy squark limit, the dominant contribution to the SI scattering comes from $t$-channel exchange of CP-even Higgs bosons~\cite{Jungman:1995df, Drees:1993bu, Drees:1992rr}, and the cross section is expressed as~\cite{Badziak:2015exr}
	\begin{equation}
	\label{equ:3-5}
        \sigma^{SI}_{\widetilde{\chi}-({\rm n})} = \frac{4\mu_{r}^2}{\pi}\left|f^{({\rm n})}\right|^2,\quad f^{({\rm n})}\approx \sum^{3}_{i=1}f^{({\rm n})}_{h_i} = \sum^{3}_{i=1}\frac{C_{h_i \widetilde{\chi}_1^0\widetilde{\chi}_1^0} C_{h_i {\rm n n}}}{2m_{h_i}^{2}},
	\end{equation}
    where $(\rm n)$ denotes nucleon, $\mu_r $ is the reduced mass of DM and nucleon, and $C_{h_i \widetilde{\chi}_1^0\widetilde{\chi}_1^0 }$($C_{h_i{\rm  n n}} $) represents the coupling of $h_i $ with DM(nucleon). Note that light Higgs boson mass appearing in the denominator of Eq.~(\eqref{equ:3-5}) can enhance the SI cross section, while on the other hand the cancellation between the contributions of different Higgs boson can suppress greatly the cross section~\cite{Badziak:2015exr}. The SD cross section is induced by the exchange of $Z$ boson, which is given by~\cite{Cao:2016cnv,Badziak:2017uto}
	\begin{eqnarray}\label{eq:sd}
	    \sigma_{\widetilde{\chi}-n/p}^{SD}/{\rm pb} \simeq C^{n/p} \times 10^{-4} \times \left(\frac{|N_{13}|^2-|N_{14}|^2}{0.1}\right)^2 \label{SD-expression}
	\end{eqnarray}
    where $n$ and $p$ denote neutron and proton respectively, $C^{p} \approx 4.0$ and $C^{n} \approx 3.1$ for the typical values of form factor $f_q^{({\rm n})}$.

    \par So far the tightest bound on the SI and SD cross sections comes from the XENON-1T experiment in 2018~\cite{Aprile:2018dbl} and the LUX measurement of DM-neutron scattering in 2017~\cite{Akerib:2017kat} respectively. Both experiments improve the limits adopted in~\cite{Cao:2016cnv} by more than six times, so we think it mandatory to update the constraints on the scenario discussed in~\cite{Cao:2016cnv} with the latest limits.
	\begin{figure}[ht]
	    \centering
	    \subfigure{\includegraphics[width=0.45\textwidth]{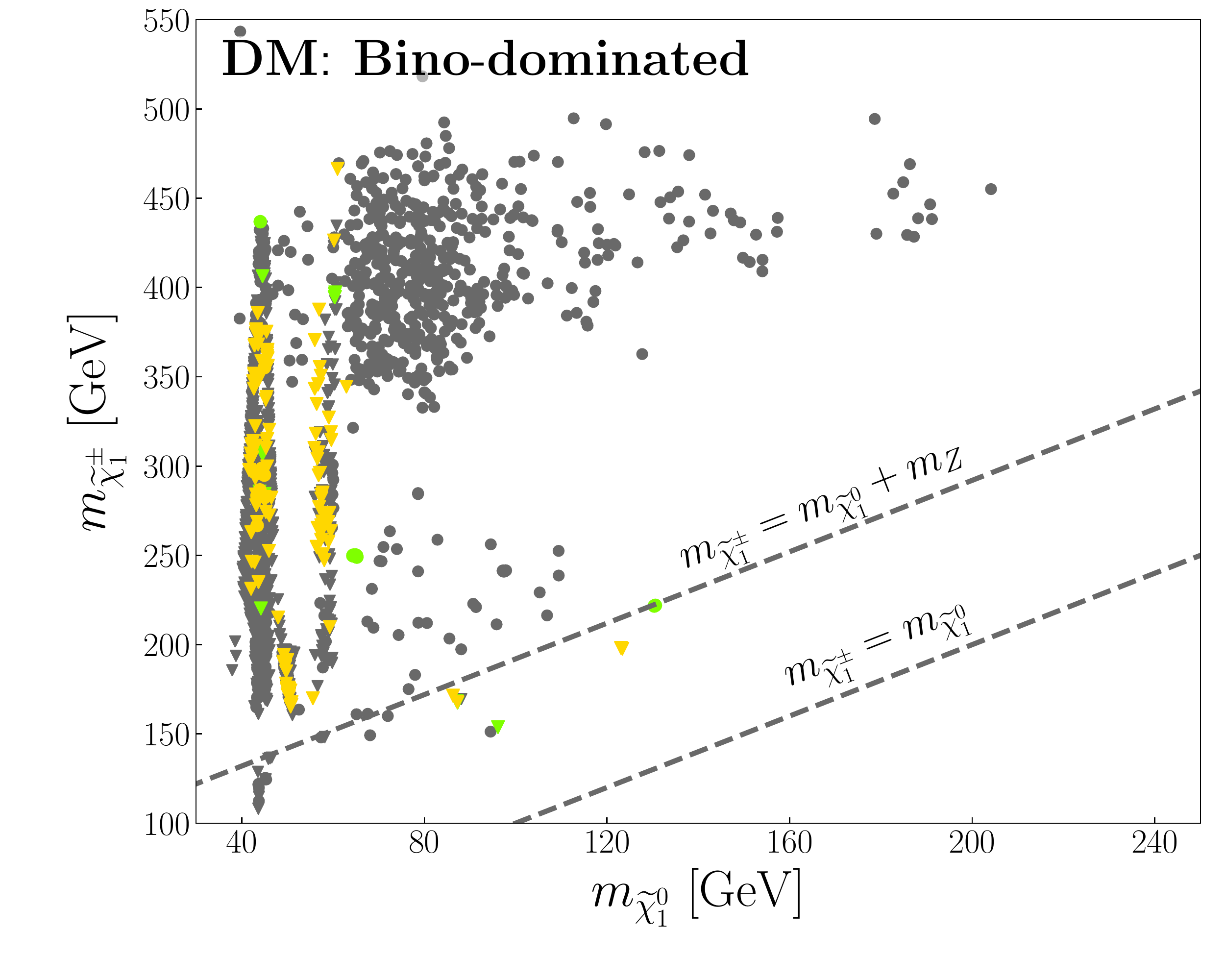}}
        \subfigure{\includegraphics[width=0.45\textwidth]{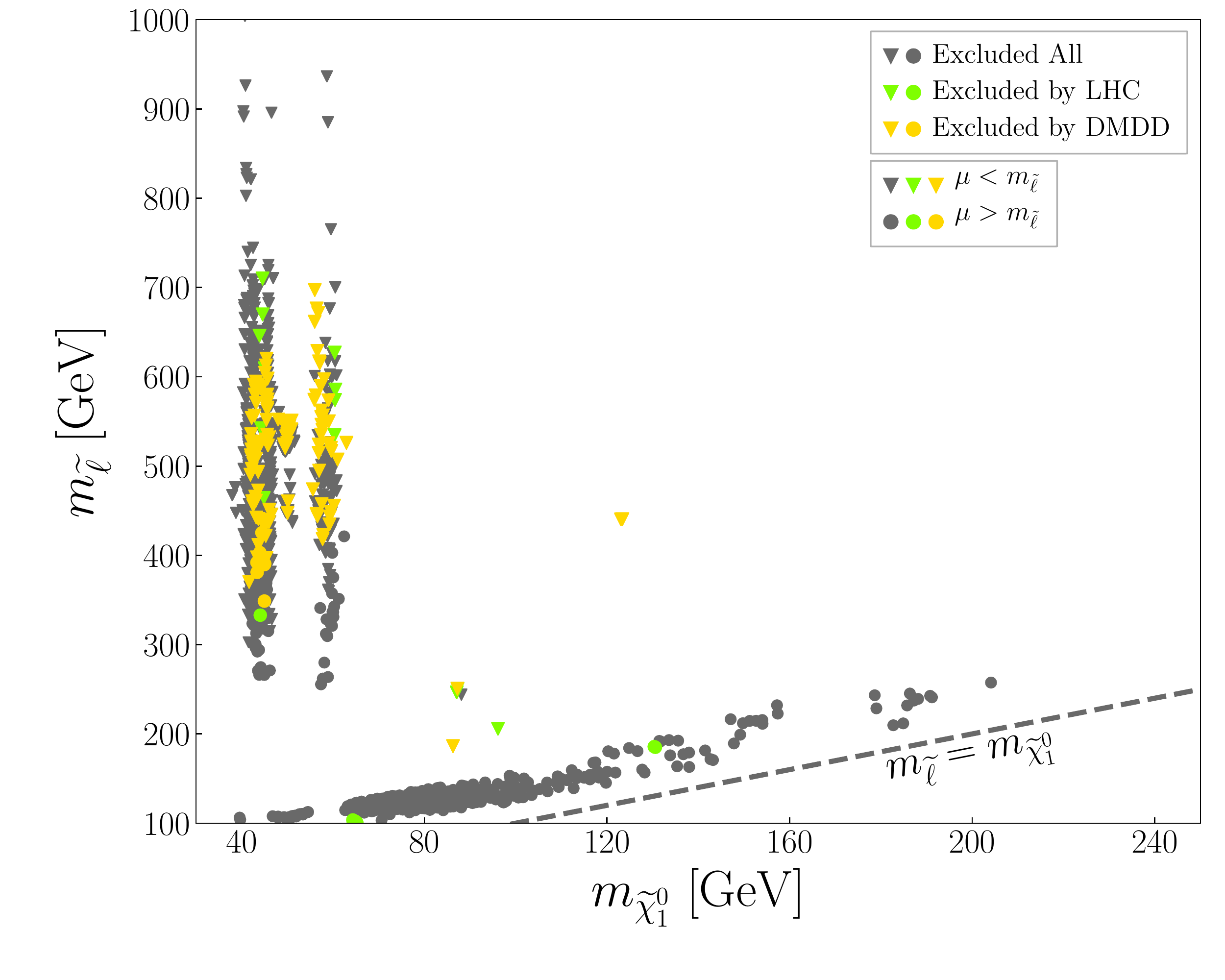}}
        \subfigure{\includegraphics[width=0.45\textwidth]{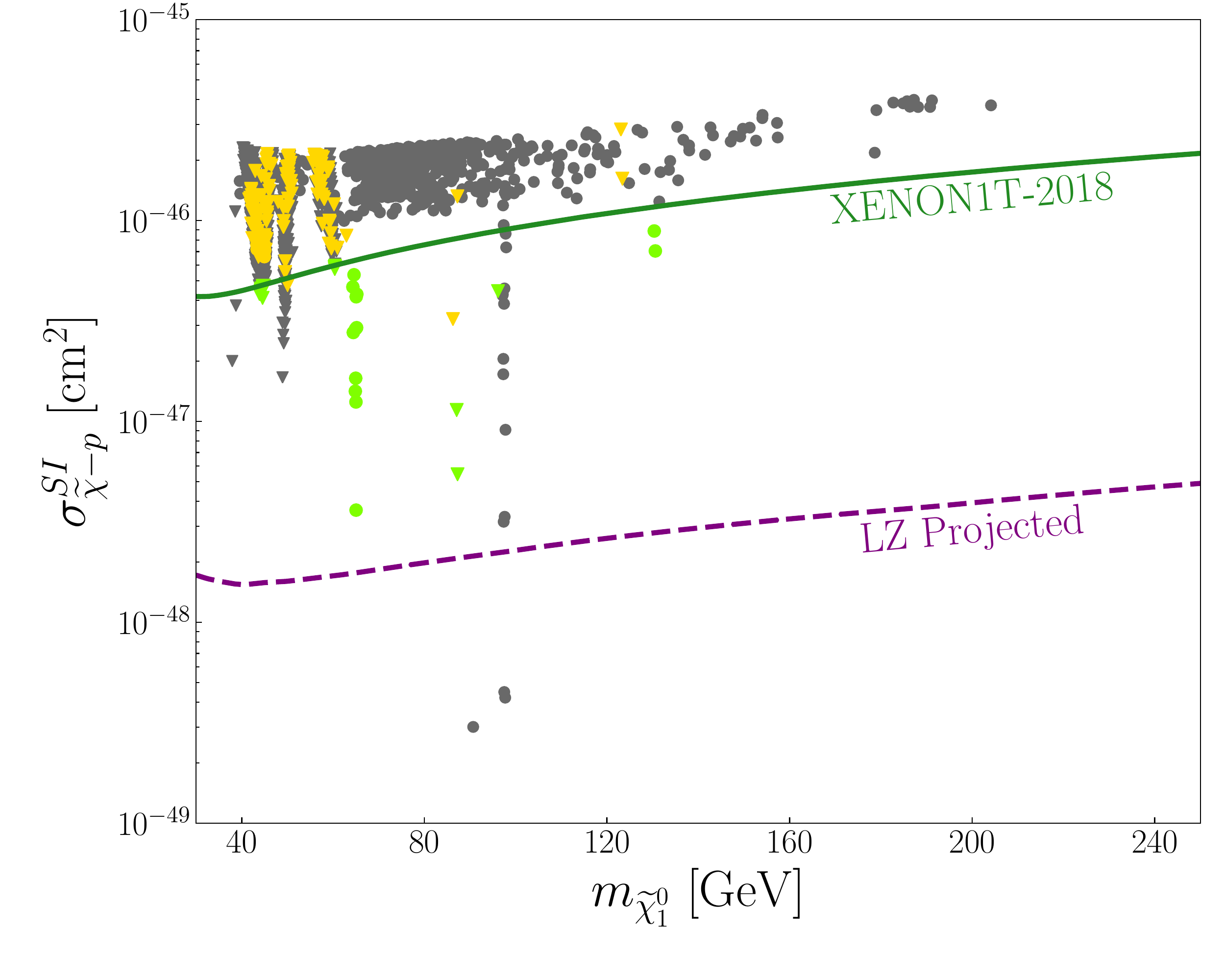}}
        \subfigure{\includegraphics[width=0.45\textwidth]{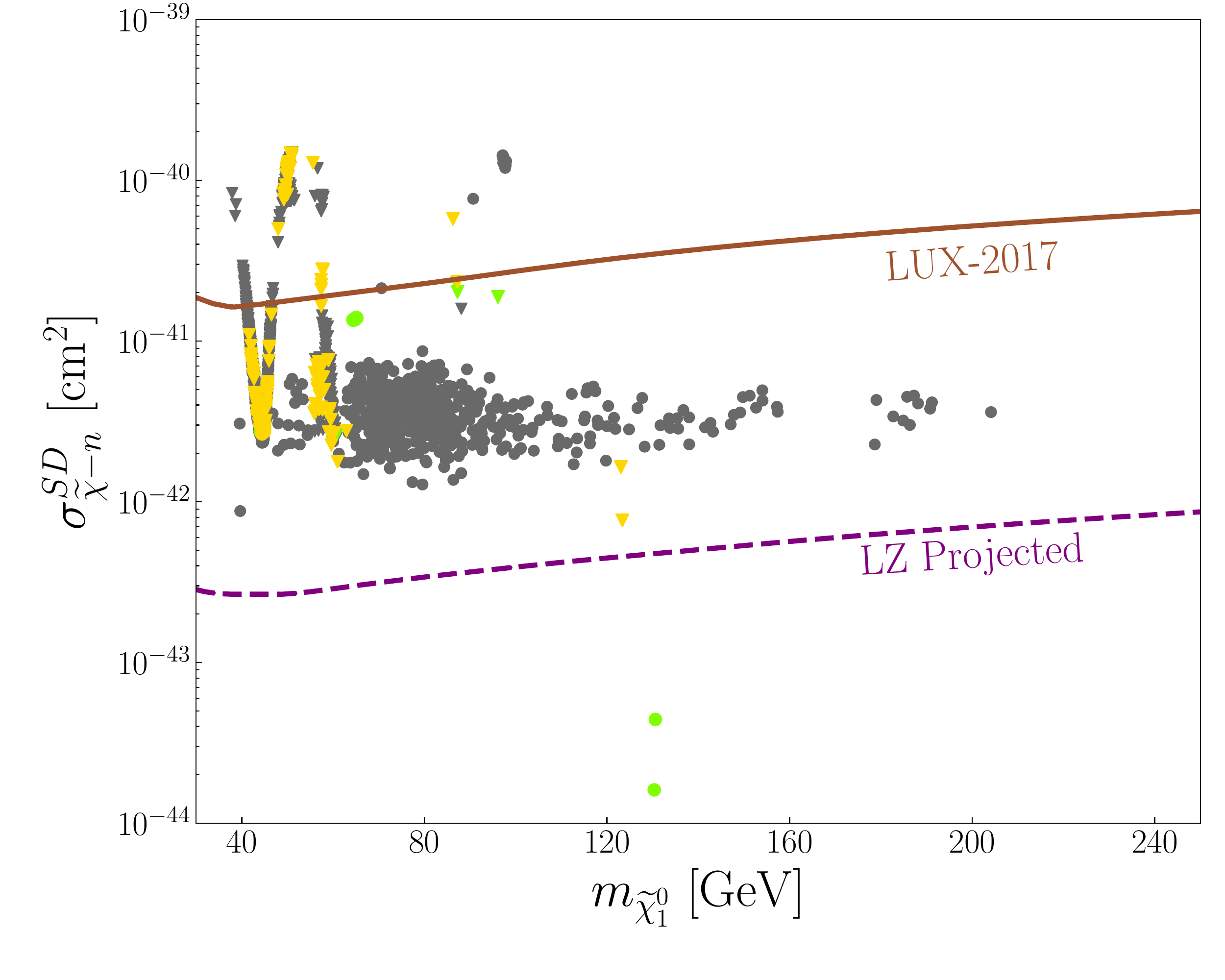}}
	    \caption{\label{fig:bino} Samples with Bino-dominated DM in the scan, which are projected on different planes with grey color indicating the points excluded by both LHC~Run~\rom{2} and DM direct detection constraints, and yellow color and green color indicating points excluded only by DM experiments and LHC experiments respectively. Samples with $m_{\widetilde{\ell}}<\mu$ and $m_{\widetilde{\ell}}>\mu$ are denoted by dot and triangle respectively.}
	\end{figure}

\subsection{\label{result13}Numerical results}
	
    Now we study the impact of the LHC experiments and the DM detection experiments on the three types of samples in natural NMSSM scenario.

    \par In Fig.~\ref{fig:bino}, we project the samples with Bino-dominated DM in the scan on $m_{\widetilde{\chi}_1^\pm}-m_{\widetilde{\chi}_1^0}$ plane (upper left panel), $m_{\widetilde{\ell}}-m_{\widetilde{\chi}_1^0}$ plane (upper right panel), $\sigma_{\widetilde{\chi}-p}^{SI}-m_{\widetilde{\chi}_1^0}$ plane (lower left panel) and $\sigma_{\widetilde{\chi}-n}^{SD}-m_{\widetilde{\chi}_1^0}$ plane (lower right panel). Most of the samples, which are marked by grey color, are excluded by both the LHC experiments and the DM detection experiments, and the rest marked by green color and yellow color are excluded only by the LHC experiments and the DM experiments respectively. Since there is no sample surviving both the constraints, it is fair to say that, at least for the assumptions made in this work, the natural NMSSM scenario with Bino-dominated DM and $\Delta_{Z/h} < 50$ is strongly disfavored by current experiments.

    \par In order to show more details about the results, we also divide the samples into two cases by different symbols: those marked by dot denote the case of $m_{\widetilde{\ell}} < \mu$, and the others marked by triangle denote the case of $m_{\widetilde{\ell}} > \mu$. The difference of the cases is that for the former case, Higgsinos prefer to decay into slepton first, which can enhance the branching ratio for leptonic final state. With the division, one can infer from Fig.~\ref{fig:bino} following facts:
    \begin{itemize}
    \item The searches for electroweakino and those for sleptons at the LHC~Run~\rom{2} are complementary to each other in excluding the samples of the natural NMSSM, which is shown by the distribution of $m_{\widetilde{\chi}_1^\pm}$ and $m_{\widetilde{\ell}}$ as a function of $m_{\widetilde{\chi}_1^0}$.

    \item For the yellow color samples, they are characterized by $\mu < m_{\widetilde{\ell}}$ and $m_{\widetilde{\chi}_1^0} \simeq m_{Z/h}/2$.  We checked that ${\rm Br}(\widetilde{\chi}_2^0 \to \widetilde{\chi}_1^0 h) > 60\%$ is slightly larger than the other parameter points in the funnel regions, which can suppress the leptonic signal of the dominant electroweakino production process  $pp \to \widetilde{\chi}_1^{\pm} \widetilde{\chi}_2^0$. The net result of these facts is that the $CLs$ values of the samples are slightly larger than 0.05, which means that they are at the edge of being excluded by the LHC analyses at $95\%$ C.L.. On the other hand, since the annihilation mechanisms set an upper bound on $\mu$ so that the coupling $C_{h \widetilde{\chi}_{1}^0 \widetilde{\chi}_{1}^0} $ is not suppressed too much, the SI cross section is moderately large, and consequently these samples are excluded by the XENON-1T experiment.
	
    \item For most green color samples, $\widetilde{\chi}_1^\pm$ is Higgsino-dominated with $m_{\widetilde{\chi}_1^\pm} \lesssim 250~{\rm GeV}$, and $h_2$ acts as the SM-like Higgs boson. In this case, the $h_2$ contribution to the SI cross section can be cancelled by the $h_1$ contribution to a great extent~\cite{Badziak:2015exr} so that the SI cross section may be as low as $10^{-48}~{\rm cm^2}$. Moreover, the SD cross section may also be suppressed by the cancellation between $|N_{13}|^2$ and $|N_{14}|^2$, which is reflected by Fig.~\ref{fig:bino} (d) and Eq.~(\ref{SD-expression}). We remind that it is actually a common case in the NMSSM with a moderately light $\mu$ that only one of the cross section is suppressed~\cite{Cao:2016cnv}, and the rare situation that both the cross sections are suppressed simultaneously was recently discussed in~\cite{Beskidt:2017xsd}.

    \item There exists some grey color samples with $m_{\widetilde{\chi}_1^0} \simeq 96~{\rm GeV}$, $m_{\widetilde{\chi}_1^\pm} \lesssim 250~{\rm GeV}$, $\sigma^{SI}_{\widetilde{\chi}-p} \lesssim 10^{-46}~{\rm cm^2}$ and meanwhile $\sigma^{SD}_{\widetilde{\chi}-n} \simeq  10^{-40}~{\rm cm^2}$. The properties of these samples are quite similar to those of the green color samples except that the cancellation between $|N_{13}|^2$ and $|N_{14}|^2$ is not strong to result in a sizable SD cross section.

    \end{itemize}
	\begin{figure}[t]
	    \centering
        \subfigure{\includegraphics[width=0.45\textwidth]{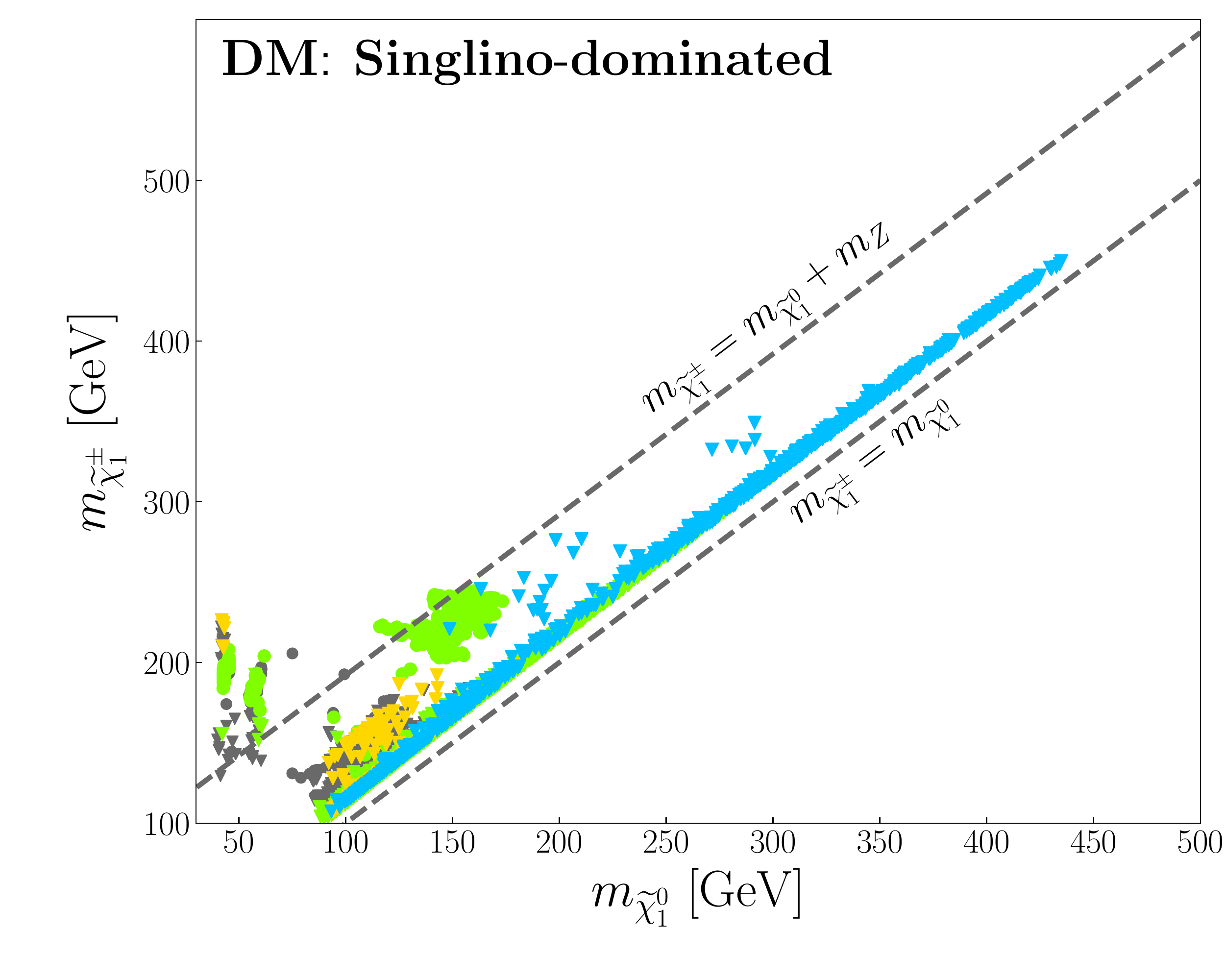}}
        \subfigure{\includegraphics[width=0.45\textwidth]{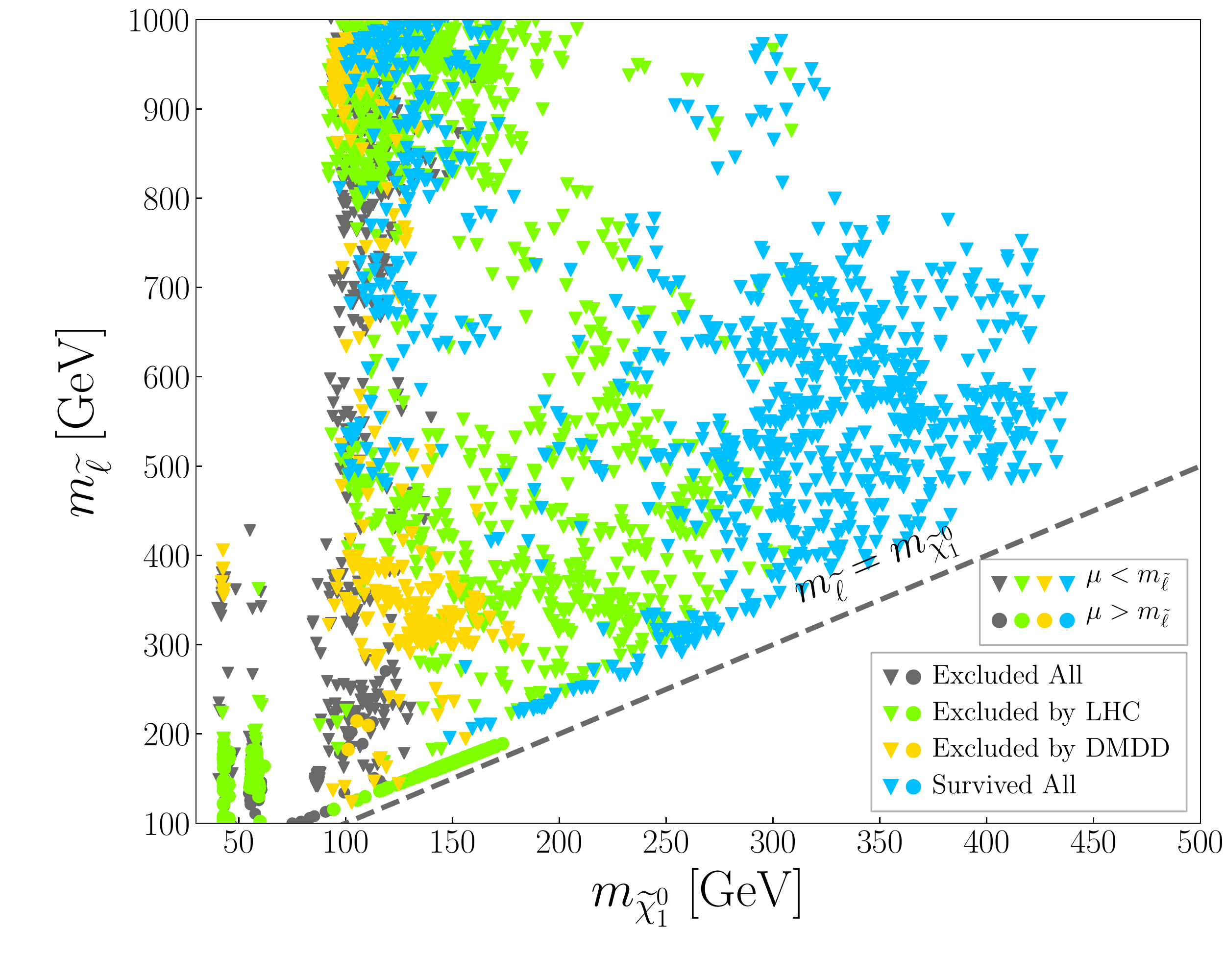}}
        \subfigure{\includegraphics[width=0.45\textwidth]{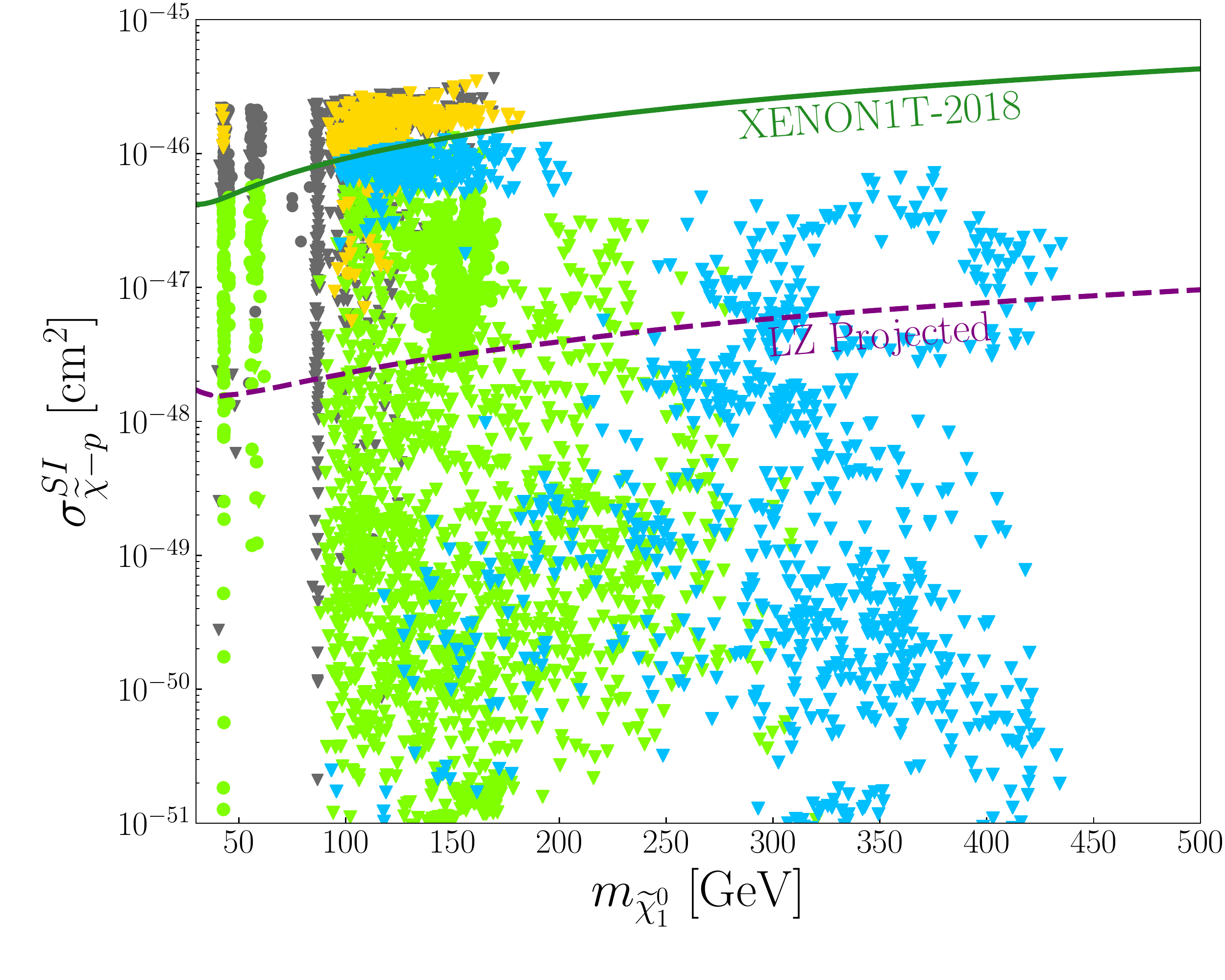}}
        \subfigure{\includegraphics[width=0.45\textwidth]{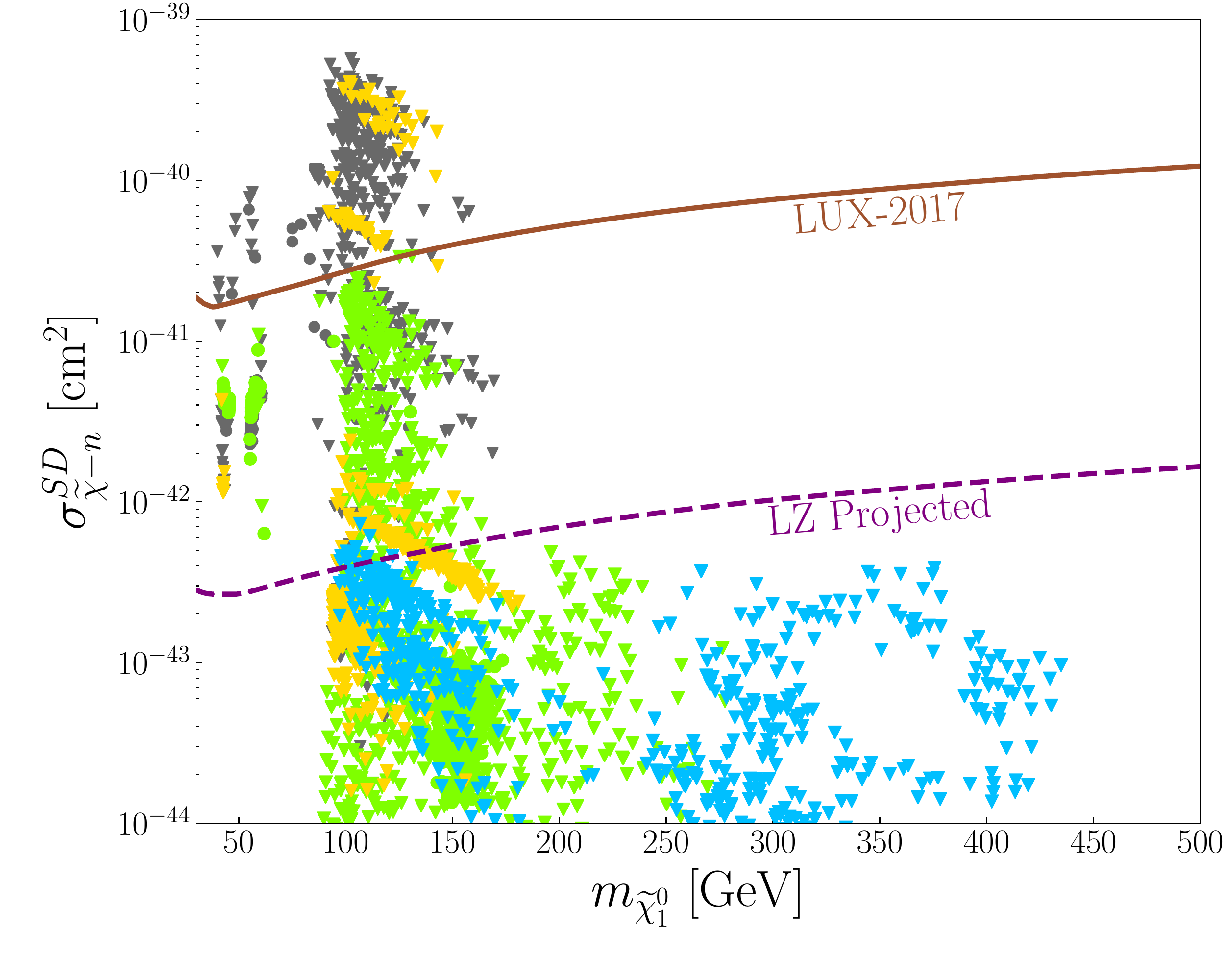}}
	    \caption{\label{fig:singlino} Same as Fig.~\ref{fig:bino}, but for Singlino-dominated DM case with blue color indicating points that survive both the LHC~Run~\rom{2} results and the DM detection results. }
	\end{figure}
	
    \par In Fig.~\ref{fig:singlino}, we illustrate the features of the Singlino-dominated DM case in a similar way to that of Fig.~\ref{fig:bino} with additional blue points standing for those which survive all the experimental constraints. From this figure, one can learn following facts:
	\begin{itemize}
	\item All the samples with $h_2$ acting as the SM-like Higgs boson satisfy $\mu \lesssim 300~{\rm GeV}$, and some of them also satisfy $M_2 \lesssim 180~{\rm GeV}$ or $M_{\widetilde{\ell}} \lesssim 400~{\rm GeV}$. While for the samples with $h_1$ corresponding to the SM-like Higgs boson, they satisfy $\mu \lesssim 450 {\rm GeV}$ with $\mu \simeq m_{\widetilde{\chi}_1^0}$ or $m_{\widetilde{\ell}} \simeq m_{\widetilde{\chi}_1^0}$. These features entail following conditions for the samples to be consistent with the experimental constraints:  moderately strong cancellation between the $h_1$ and $h_2$ contributions to the SI cross section, $|N_{13}|^2 \simeq |N_{14}|^2$ as well as the suppressed spectrum of the sparticles with $\widetilde{\chi}_1^0$~\cite{Cao:2016nix,Cao:2016cnv}\footnote{As was discussed in numerous literature, the final states of neutralino/chargino pair production in this case become soft to be indistinguishable from SM background processes at LHC.}.
	
	\item Similar to the Bino-dominated DM case, samples featured by $m_{\widetilde{\chi}_1^0} \simeq m_{Z/h}/2$ or $\mu > m_{\widetilde{\ell}}$ are completely excluded by the current experimental limits. Constrains from the LHC electroweakino searches play critical roles.
	
    \item Nearly all the samples with an approximate degeneracy of Wino-like $\widetilde{\chi}_1^\pm$ with $\widetilde{\chi}_1^0$ in mass are excluded. Some of them may survive the constraints from the LHC experiments and the XENON-1T experiment, but are excluded by the measurement on the SD cross section. These samples correspond to the yellow color samples in the four panels of Fig.~\ref{fig:singlino} featured by $ 90~{\rm GeV} \lesssim m_{\widetilde{\chi}_1^0} \lesssim 120~{\rm GeV}$, $m_{\widetilde{\chi}_1^\pm} \lesssim 160~{\rm GeV}$, $m_{\widetilde{\ell}} \gtrsim 400~{\rm GeV}$, $\sigma^{SI}_{\widetilde{\chi}-p} \lesssim 10^{-46}~{\rm cm^2}$ and $\sigma^{SD}_{\widetilde{\chi}-n} \gtrsim 2 \times 10^{-41}~{\rm cm^2}$.

    \item The samples satisfying $ m_{\widetilde{\chi}_1^0} \simeq \mu$ and forming a line parallel to $m_{\widetilde{\chi}_1^\pm} = m_{\widetilde{\chi}_1^0}$ are more complicated. For yellow samples featured with $ 90~{\rm GeV} \lesssim m_{\widetilde{\chi}_1^0} \lesssim 200~{\rm GeV}$ in Fig.~\ref{fig:singlino} (a) and $m_{\widetilde{\ell}} < 400~{\rm GeV}$ in Fig.~\ref{fig:singlino} (b), the LHC experiments have no exclusion capability. The SI cross section is sizable in comparison its detection limit, which varies from $2 \times 10^{-47}~{\rm cm^2}$ to $3 \times 10^{-46}~{\rm cm^2}$, while the SD cross section is suppressed too much to be less than $2 \times 10^{-42}~{\rm cm^2} $.  For the green samples, they satisfy $\mu < m_{\widetilde{\ell}}$, and the constraints of the LHC~Run~\rom{2} mainly come from the associated production of Wino-like chargino and neutralino.

    \item Most important, there exist samples that survive all the constraints, which correspond to the co-annihilation region of the DM with Higgsinos to get the right relic density, and are marked by blue color in Fig.~\ref{fig:singlino}. In addition, some of these samples may also co-annihilate with slepton, and consequently the mass splitting between the DM and the Higgsinos can be slightly larger in getting the right DM relic density. Compared with the green samples discussed above, sleptons and Wino-like neutralino/chargino are heavier to escape the constraints from LHC~Run~\rom{2}. Moreover, we note that there are surviving samples with high singlet purity ($N_{15}^2 > 0.99$). In this case, the DM decouples with SM particles so that both SI and SD cross sections are lower than the future LZ detection limits \cite{Akerib:2018lyp}. This case was recently emphasized in~\cite{Beskidt:2017xsd}. For the other samples without such high singlet purity, the SI cross section may be at the order of $10^{-47}~{\rm cm^2}$, which will be explored by near future DM direction detection experiments, and the SD cross section is usually less than $5 \times 10^{-43}~{\rm cm^2}$ which is far below its current detection limits.

	\end{itemize}
	
    \begin{table}[t]
        \centering
        \resizebox{\textwidth}{!}{
        \begin{tabular}{cccccccccccccc}
        \hline
                    & $m_{\widetilde{\chi}_1^0}$
                    & $m_{\widetilde{\chi}_1^\pm}$
                    & $M_{1}$
                    & $M_{2}$
                    & $\mu$
                    & $\Omega h^2$
                    & $\sigma_{\widetilde{\chi}-p}^{SI}\ (\rm{cm}^2)$
                    & $\sigma_{\widetilde{\chi}-n}^{SD}\ (\rm{cm}^2)$
                    & $\Delta_Z$
                    & $\Delta_h$\\
        \hline
        \texttt{P1} & 94.9  & 141.4 & 498.0 & 165.1 & 231.2 & 0.12266   & \SI{4.65e-50} & \SI{6.00e-41}   & 20.9  & 42.8 \\
        \texttt{P2} & 119.1 & 133.7 & 684.2 & 1021.9    & 131.6 & 0.12488   & \SI{6.91e-47}   & \SI{3.34e-43}   & 28.1  & 20.4 \\
        \hline
        \end{tabular}}
        \caption{Detailed information about two benchmark points \texttt{P1} and \texttt{P2} for Singlino-dominated DM case. All quantities with mass dimension are given in units of GeV.}\label{tab:bk}
    \end{table}

    \par In order to emphasize the property of the samples with Wino dominated $\widetilde{\chi}_1^\pm$ and Higgsino dominated $\widetilde{\chi}_1^\pm$ in the co-annihilation region, we choose two benchmark points \texttt{P1} and \texttt{P2} with their detailed information presented in Tab.~\ref{tab:bk}. Both the points pass the LHC constraints, but their behaviors confronted the DM detection limits are quite different: for the Wino dominated $\widetilde{\chi}_1^\pm$ case (point \texttt{P1}), the SI cross section is far below its detection limit while the SD cross section is around its detection limit, and the situation is reversed for the Higgsino dominated $\widetilde{\chi}_1^\pm$ case (point \texttt{P2}).
	\begin{figure}[ht]
	    \centering
	    \subfigure{\includegraphics[width=0.45\textwidth]{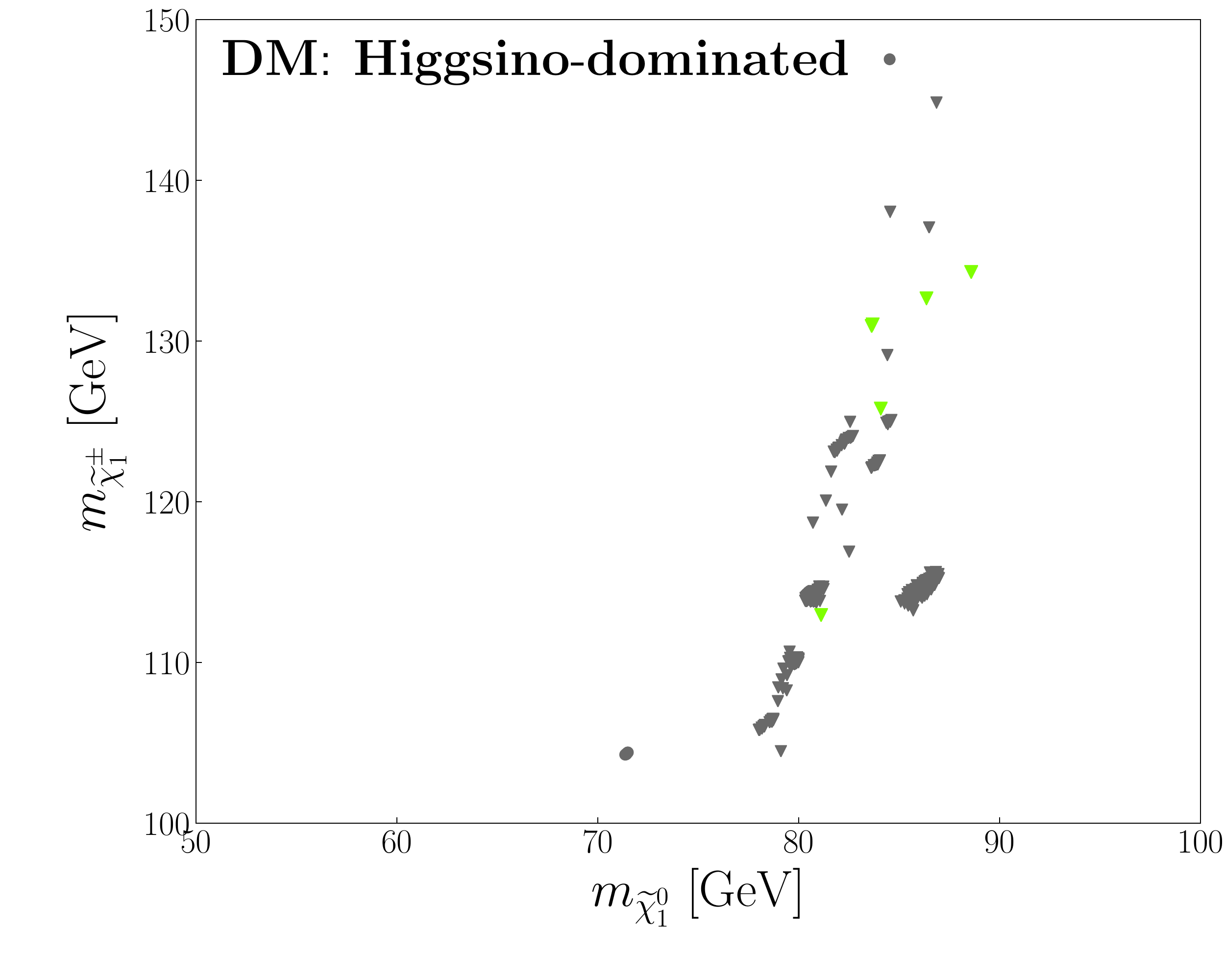}}
        \subfigure{\includegraphics[width=0.45\textwidth]{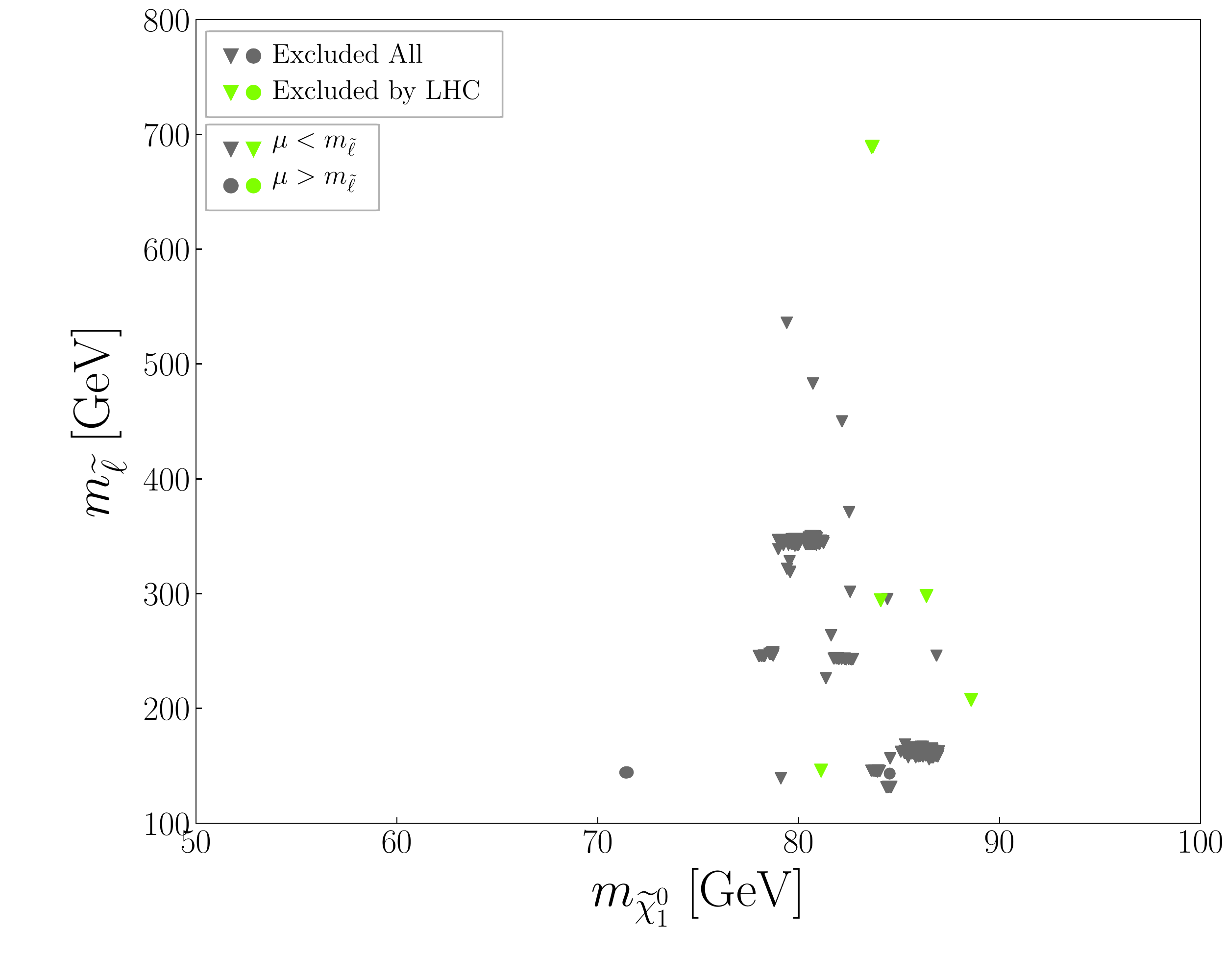}}
        \subfigure{\includegraphics[width=0.45\textwidth]{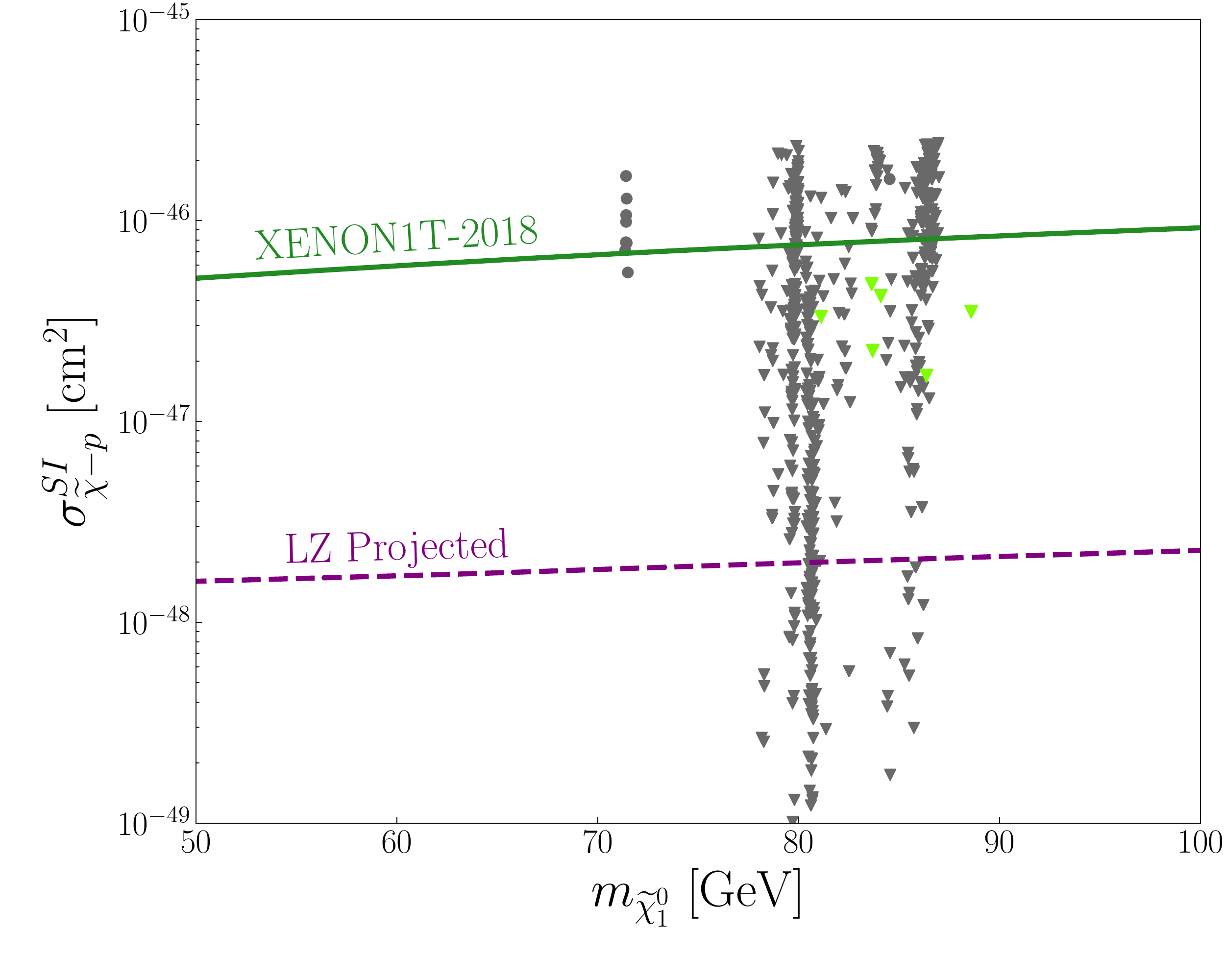}}
        \subfigure{\includegraphics[width=0.45\textwidth]{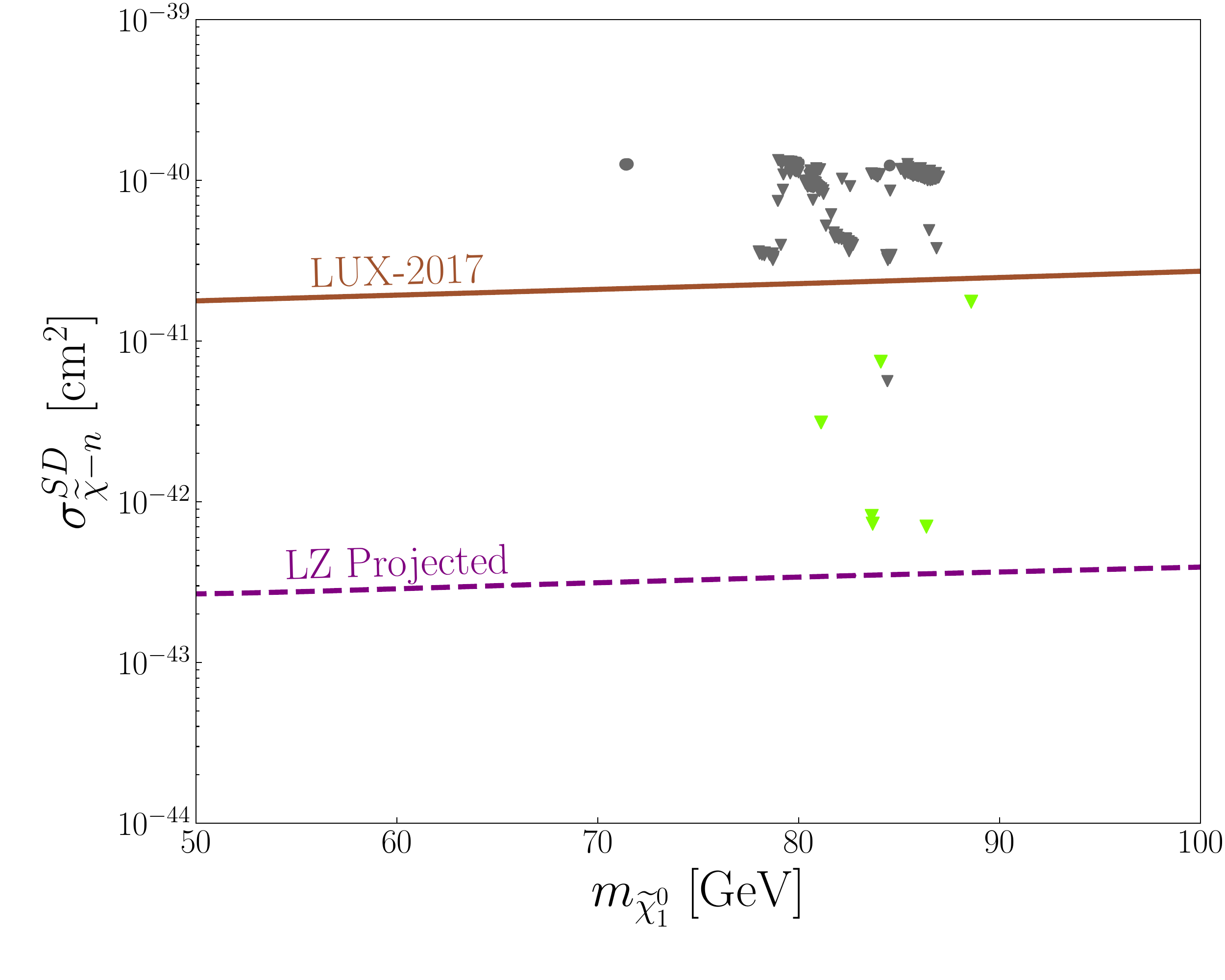}}
	    \caption{\label{fig:higgsino} Similar to Fig.~\ref{fig:bino}, but for Higgsino-dominated DM case. }
	\end{figure}
    \par Finally we consider the Higgsino-dominated DM case. This kind of samples predict a light CP-even Higgs boson with $m_{h_1} < 125~{\rm GeV}$, $ 70~{\rm GeV} \lesssim m_{\widetilde{\chi}_1^0} \lesssim 100~{\rm GeV}$, $\mu \lesssim 160~{\rm GeV}$ and moderately large mixing between Higgsino and Singlino in forming neutralino mass eigenstates~\cite{Cao:2016nix}. In Fig.~\ref{fig:higgsino}, we project the samples on different planes like what we did in Fig.~\ref{fig:bino}. From this figure, one can learn following facts:
    \begin{itemize}
    \item Although the mass splittings between $\widetilde{\chi}_{2/3}^0/\widetilde{\chi}_{1}^\pm$ and $\widetilde{\chi}_{1}^0$ are relatively small, the quite large cross section of neutralino/chargino pair production leads to the exclusion of all the samples by the electroweakino searches described in Sec.~\ref{sec:lhc}. Some of the samples can also be excluded by the slepton searches at LHC.
    \item The XENON-1T experiment can only exclude a small portion of the samples due to the strong cancellation of the contributions of $h_1$ and $h_2$ to the SI cross section, while the LUX-2017 limits on the SD cross section are rather effective in excluding the samples. Consequently, few samples are allowed by DM direct detection experiments.

    \end{itemize}
	
	\begin{figure}[ht]
	\centering
	\includegraphics[width=0.49\textwidth]{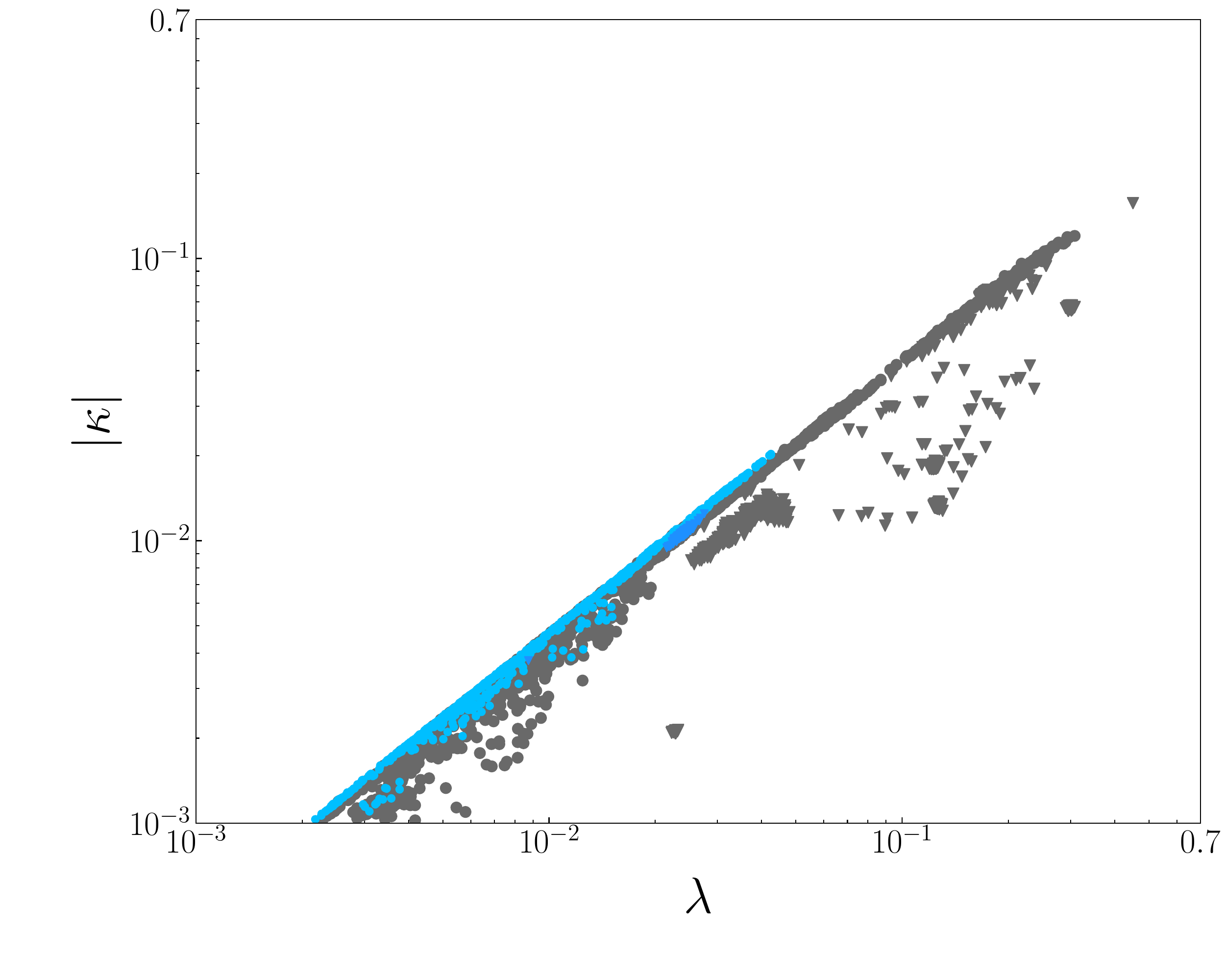}
	\includegraphics[width=0.49\textwidth]{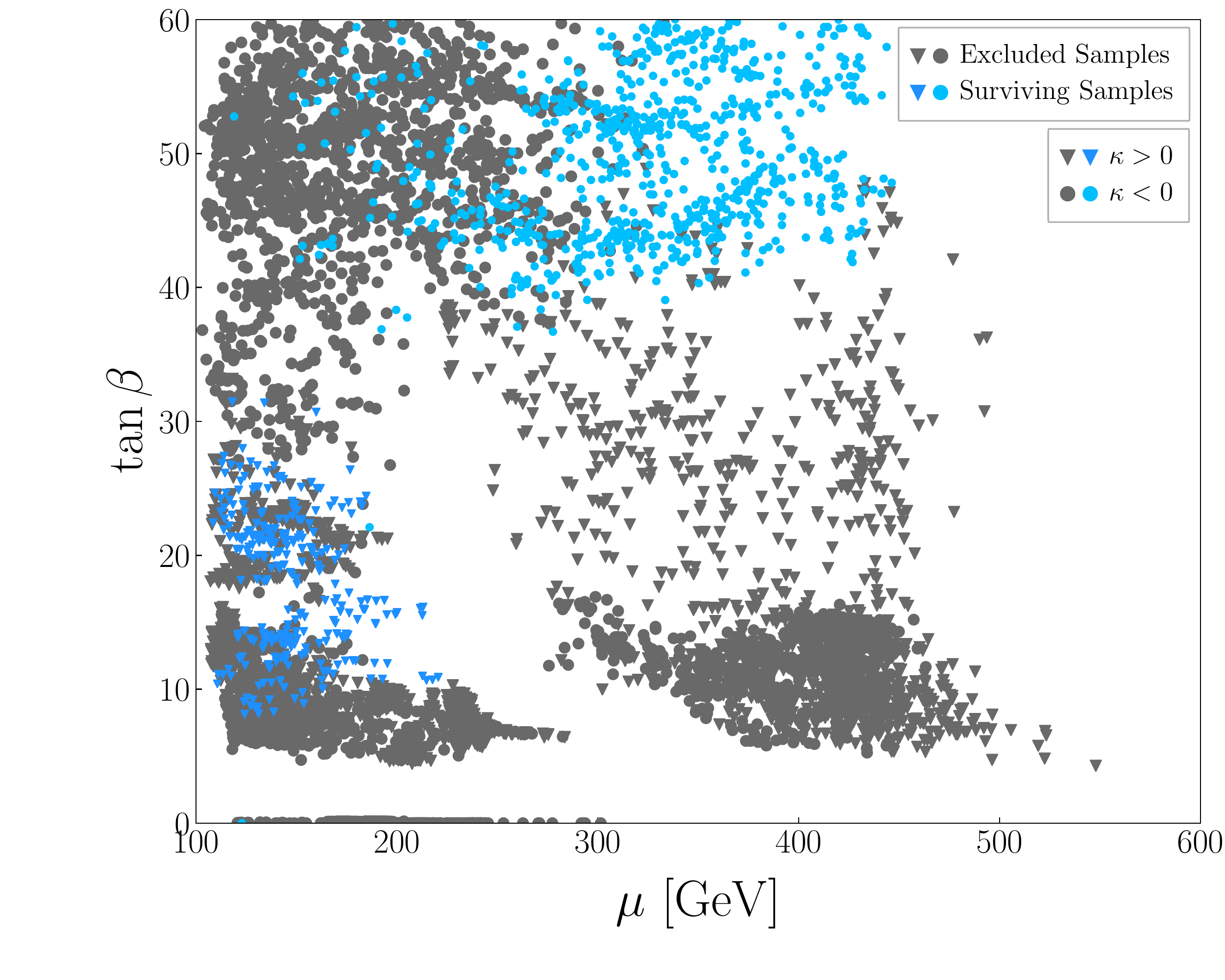}
	\caption{\label{fig:para}Samples obtained in the scan, which are projected on $\lambda-\kappa$ and $\mu-\tan{\beta}$ planes. The grey color samples have been excluded by the LHC~Run~\rom{2} experiments and the DM direct detection experiments, and the blue ones are still experimentally allowed. Samples with $\kappa > 0$ and $\kappa < 0$ are denoted by triangle and dot respectively.}
	\end{figure}

	\begin{figure}[ht]
	\centering
	\includegraphics[width=0.95\textwidth]{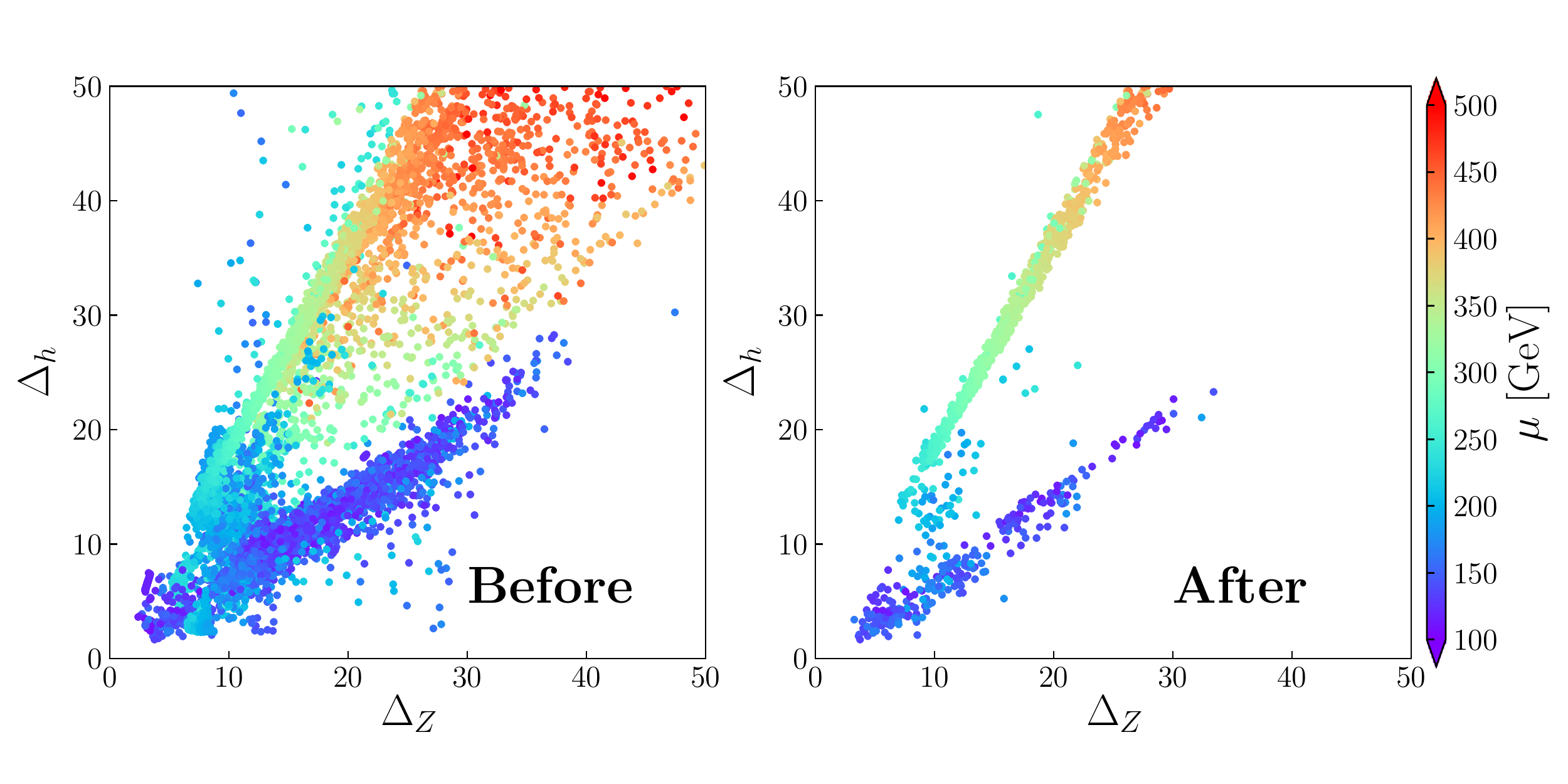}
	\caption{\label{fig:ft} Fine tunings of the natural NMSSM scenario before and after considering the LHC~Run~\rom{2} and DM detection results with different colors representing the values of $\mu$, which is indicated by the color bar on the right side of the figure.}
	\end{figure}

	\begin{figure}[ht]
	\centering
	\includegraphics[width=0.49\textwidth]{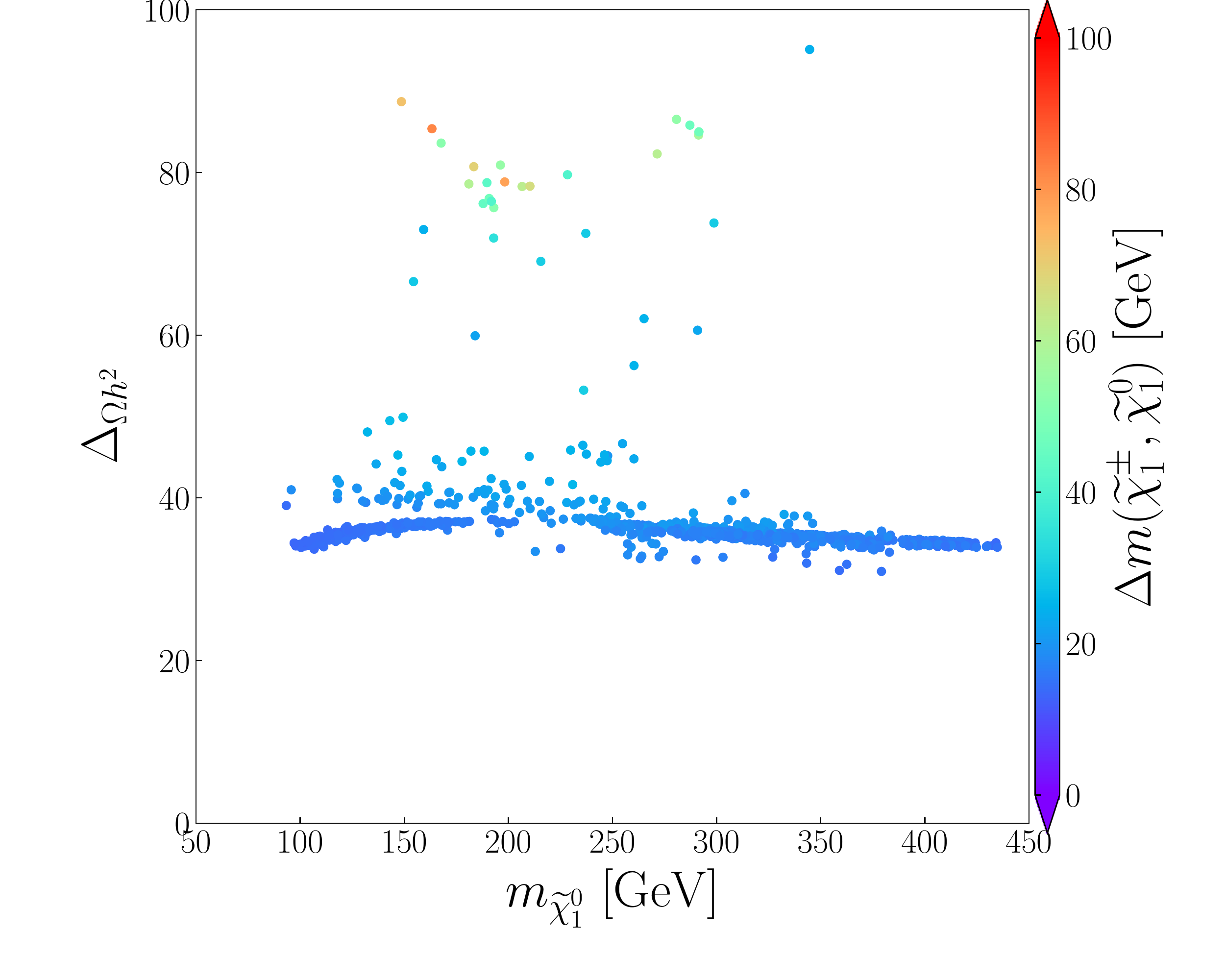}
	\includegraphics[width=0.49\textwidth]{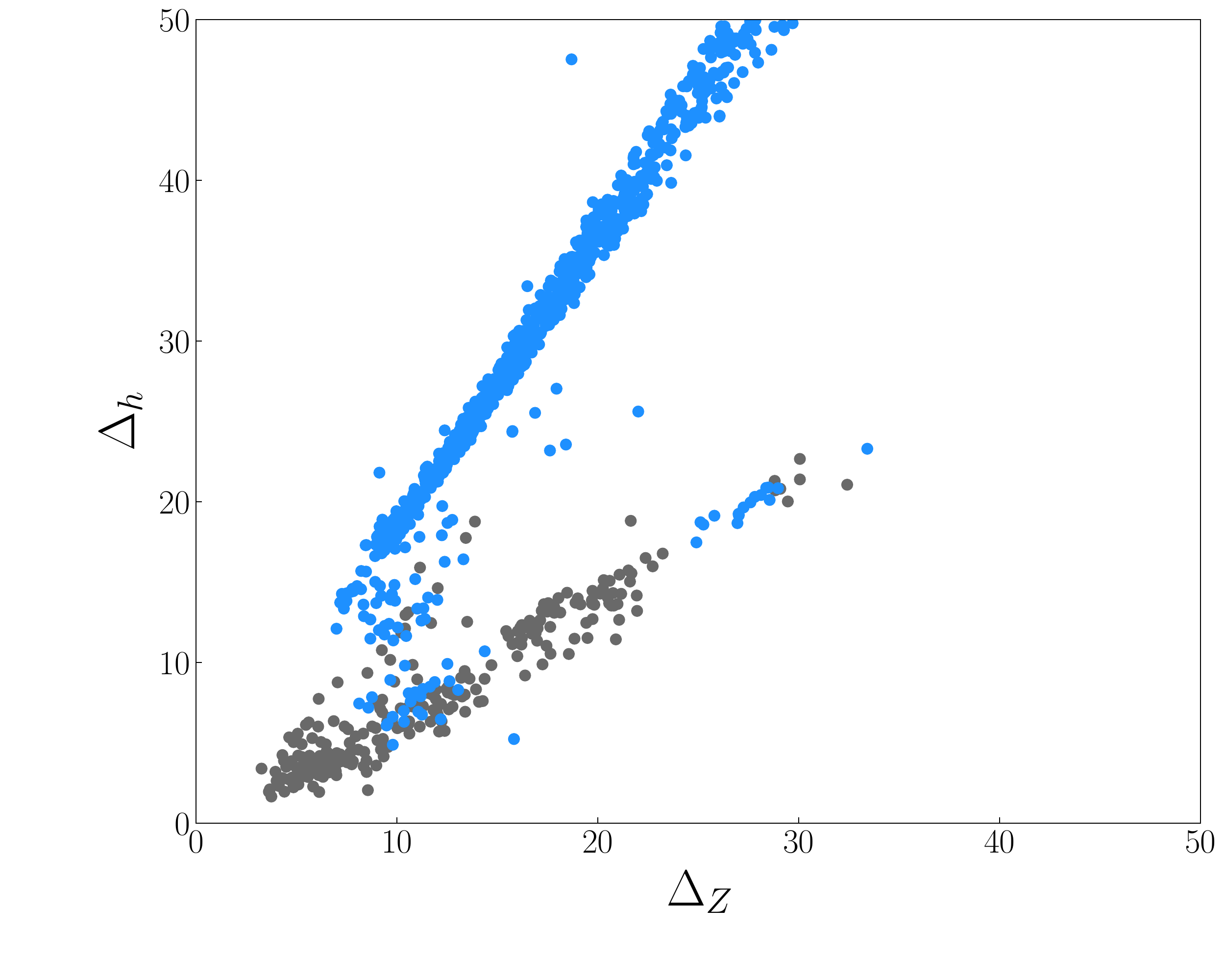}
	\caption{\label{fig:muon} Samples in the scan Eq.~(\ref{eq:input}) surviving all the constraints considered in this work, which are projected on $m_{\widetilde{\chi}_1^0}-\Delta_{\Omega h^2}$ plane and $\Delta_{Z}-\Delta_{h}$ plane, respectively. In the left panel, the color indicates the mass difference between $\widetilde{\chi}_1^\pm$ and $\widetilde{\chi}_1^0$, and in the right panel, the blue (gray) points stand for the samples which are able to (unable to) explain the discrepancy of the muon anomalous magnetic moment at $2\sigma$ level. }
	\end{figure}

    \section{\label{sec:status}Status of the natural NMSSM}

    The results in previous sections reveals that in the natural NMSSM scenario with $\Delta_{Z/h} \leq 50$, only the  Singlino-dominated DM case can survive the tight experimental constraints if the correlation $\mu \simeq m_{\widetilde{\chi}_1^0}$ holds. This has non-trivial impact on the parameter space of the NMSSM and also on the fine tunings of the theory. In Fig.~\ref{fig:para}, we project the samples obtained in the scan on $\lambda-\kappa$ plane and $\mu-\tan{\beta}$ plane with the grey (blue) color samples being experimentally excluded (allowed). This figure indicates that, after considering the constraints, the scenario is restricted in certain narrow corners of the NMSSM parameter space, which is featured by $\lambda/\kappa \simeq 2.5$ with $\lambda \approx 0.02$, $100~{\rm GeV} \lesssim \mu \lesssim 200~{\rm GeV}$ and $8 \lesssim \tan{\beta} \lesssim 32$ for $\kappa > 0$ and by $\lambda/\kappa \simeq -2.5$ with $\lambda \lesssim 0.05$, $\mu \lesssim 460~{\rm GeV}$ and $36 \lesssim \tan{\beta} < 60$ for $\kappa < 0$. In Fig.~\ref{fig:ft}, we show the fine tuning indicators of the scenario before and after considering the LHC~Run~\rom{2} and DM detection results with different colors representing the values of $\mu$ (see the color bar on the right side of the figure). This figure shows again that the experimental constraints are very powerful in limiting the scenario and have reduced significantly the range of $\Delta_{Z/h}$.

    \par Given the status of the natural NMSSM, it is interesting to ask following questions:
    \begin{enumerate}
        \item Since the DM relic density is another precisely measured quantity, what is the tuning needed to get its measured value?
        \item Is the natural NMSSM scenario able to explain the discrepancy of muon anomalous magnetic moment?
        \item What are the effects on the conclusions given above if one relaxes the requirement on the fine tuning measurements by $\Delta_Z, \Delta_h \leq 100$?
        \item  What will happen if one takes the value of the DM relic density measured by Planck just as an upper bound?
    \end{enumerate}

    \par In order to answer the first question, we define the fine tuning measurement of the density as
    \begin{equation}
	    \Delta_{\Omega h^2}\equiv \max_{i}\left|\frac{\partial\log \Omega h^2}{\partial\log p_i}\right|,
	\end{equation}
    where $p_i$ denotes the variables in Eq.~(\ref{eq:input}), and present $\Delta_{\Omega h^2}$ for the surviving samples on $m_{\widetilde{\chi}_1^0}-\Delta_{\Omega h^2}$ plane in Fig.~\ref{fig:muon} with the color bar denoting the mass splitting between $m_{\widetilde{\chi}_1^\pm}$ and $m_{\widetilde{\chi}_1^0}$.  This panel indicates that for the samples in the co-annihilation region of the Singlino-dominated DM with Higgsinos, $\Delta_{\Omega h^2} \simeq 35$ which is insensitive to the DM mass, while for those in the co-annihilation region with sleptons, $\Delta_{\Omega h^2}$ can be as large as 95. We note that our results about $\Delta_{\Omega h^2}$ coincide with those in~\cite{Cao:2019aam}.  As for the second question, we categorize the surviving samples by whether they can explain the muon $g-2$ anomaly at $2\sigma$ level or not, and present them on the $\Delta_{Z}-\Delta_{h}$ plane of Fig.~\ref{fig:muon}. The samples marked by blue color are able to explain the anomaly, while those marked by grey color fail to do so. This panel indicates that the explanation of the anomaly places additional restrictions on the scenario, and consequently, due to the shrink of the allowed parameter space from  $\mu \gtrsim 100~{\rm GeV}$ to $\mu \gtrsim 150~{\rm GeV}$, the lower bound on $\Delta_Z$ ($\Delta_h$) is shifted from  2(2) to 7(5). In getting the results, we use the default setting of the \texttt{NMSSMTools} to take into account the theoretical and experimental uncertainties of the anomaly.
    With respect to the third question, we note that relaxing the constraint on $\Delta_Z$ and $\Delta_h$ will allow the parameter $\mu$ to vary over a broader range since both fine tuning indicators are sensitive to $\mu$. A larger $\mu$ can suppress the rate of electroweakino pair production at the LHC as well as the DM-nucleon scattering rate, which is helpful for the theory to escape the experimental constraints. Our results from an additional scan of the parameter space in Eq.~(\ref{eq:input}) indicate that allowing $\Delta_{Z/h} \leq 100$ can increase the samples in the slepton co-annihilation region for the Bino-dominated DM case and the Higgsino co-annihilation region for the Singlino-dominated DM case without violating the constraints. The results also show that the Higgsino-dominated DM case is scarcely affected by relaxing the fine tuning measurements. Finally, we point out that taking a lower value of the density $\Omega^\prime h^2$ is equivalent to relax the upper bound on the cross section of the DM-nucleon scattering by a factor $(\Omega^\prime h^2)/0.1187$, and consequently the constraints from the DM detection experiments are weakened. As far as the Bino-dominated DM case and the Singlino-dominated DM case are concerned, a lower relic density can be achieved by narrowing the mass gap between the DM and its co-annihilating particles. This will not affect the constraints from the LHC experiments. Moreover, without the right relic density, the DM may be a pure Higgsino particle. In this case, its relic density is less than $0.01$~\cite{Baer:2012uy,Baer:2013gva}, and its scattering with nucleon is suppressed greatly since there is no triple doublet-Higgs interaction in the superpotential of the NMSSM.

    \par Before we end this section, we have following comments about our results:
    \begin{itemize}
    \item In our discussion, we do not consider the constraints from the direct search for top squarks at the LHC~Run~\rom{2}~\cite{Aaboud:2017ayj}. We checked that the surviving samples in Fig.~\ref{fig:para} may predict the lighter stop mass as low as about $600~{\rm GeV}$, and part of those samples are sure to be tested by the search. This will further shrink the parameter space of the scenario.
    \item We note that the discovery potential for the electroweakino production process $p p \to \widetilde{\chi}_1^{\pm}\widetilde{\chi}_2^0$ at future high luminosity LHC has been estimated by ATLAS collaboration~\cite{ATL-PHYS-PUB-2015-032, ATL-PHYS-PUB-2014-010} and CMS collaboration~\cite{CMS-PAS-SUS-14-012} in trilepton and WH channels. Using the relevant analysis codes for ATLAS collaboration at 14 TeV LHC~\cite{ATL-PHYS-PUB-2014-010},  which was provided by the package \texttt{CheckMATE}, we find the analysis has no exclusion capability for the surviving samples even for the luminosity as high as $3000~\rm{fb}^{-1}$.
    \item As we mentioned before, in order to satisfy the strong constraint of XENON-1T experiment on the SI cross section, the $h_2$ contribution must be cancelled greatly by the $h_1$ contribution. This induces another kind of fine tuning in DM physics which is different from the tuning in the electroweak symmetry breaking and was discussed in~\cite{Perelstein:2012qg}. The origin of the tuning comes from two aspects. One is that the Higgsino mass $\mu$ should be of ${\cal{O}}(10^2~{\rm GeV})$ to predict $m_Z$ in a natural way. Such a light $\mu$ can enhance the SI cross section greatly. The other is that the parameters in the Higgs sector have been tightly limited by the LHC search for Higgs bosons, and this determines the relative size of each $h_i$ contribution to the cross section~\cite{Badziak:2015exr}. Take the heavy doublet dominated Higgs boson as an example, its contribution to the SI cross section can be neglected safely in most cases since the search for extra Higgs bosons at the LHC has required its mass at TeV scale, which can suppress the contribution greatly.

    \end{itemize}

    \section{Summary}\label{sec:conclusion}
	
    In this work, we explore the constraints from the direct searches for electroweakino and slepton at the LHC~Run~\rom{2} and the latest DM direct detection experiments on the natural NMSSM scenario for three types of samples, namely those with Bino, Singlino and Higgsino as dominant DM component respectively. We have following observations:
    \begin{itemize}
    \item Moderately light Higgsinos are favored by this scenario, which usually results in detectable leptonic signals at the LHC as well as large DM-nucleon scattering rate. Moreover, in some cases Wino and sleptons with mass around several hundred GeVs are also predicted. This situation makes the scenario to be testable readily by the experiments, and surviving these experiments necessitates great cancellation among different Higgs contributions to the SI cross section of DM-nucleon scattering, $|N_{13}|^2 \simeq |N_{14}|^2$ and suppressed sparticle spectrum. This, on the other hand, induces a kind of tuning of the theory which is other than the fine tuning in electroweak symmetry breaking sector.
    \item The signal of the electroweakino/slepton pair productions at the LHC~Run~\rom{2} and the SI and SD cross section for DM-nucleon scattering are sensitive to different parameter space of the NMSSM, and their constraints are complementary to each other in excluding the samples of the natural NMSSM scenario. As far as each kind of the experiments is concerned, its individual constraint is strong enough to exclude most samples of the scenario.
    \item With the assumptions made in this work, the samples with Bino- or Higgsino-dominated DM are completely excluded by the experiments, and most samples for Singlino-dominated DM case are also excluded. As a result, some input parameters of the natural NMSSM scenario are restricted in certain narrow corners of the NMSSM parameter space.
    \item Although future LHC experiments and DM detection experiments can further limit the parameter space of the natural NMSSM scenario, there exist special parameter regions where the Singlino-dominated DM decouples from the SM sector. In this case, neither LHC experiments nor DM direct detection experiments can probe the scenario.

    \end{itemize}

    \par In summary, given the tight experimental constraints on the natural NMSSM scenario, its charm is fading, and one may either accept the current situation of the theory or insist on the fine tuning criteria as a guidance of new physics to construct more elaborated theories. For the latter choice, the seesaw extensions of the NMSSM, which is motivated by neutrino mass, provide an economical solution to the problem of the strong constraints by choosing the lightest sneutrino as the DM candidate~\cite{Cao:2018iyk,Cao:2019aam,Cao:2017cjf}. As was shown in~\cite{Cao:2018iyk,Cao:2019aam}, a moderately light $\mu$ in this framework is favored not only by predicting naturally $Z$ boson mass, but also by predicting right DM physics. The signals of sparticles at the LHC may be quite different from those in the MSSM or NMSSM, which is helpful to evade collider constraints~\cite{Cao:2018iyk,Cao:2017cjf}.

	\section* {\sc Acknowledgement}
	Pengxuan Zhu would like to thank Prof. Rob Hardy and Prof. Peng Tian for their patience and guidance in academic writing. This work is supported by the National Natural Science Foundation of China (NNSFC) under grant No. 11575053, 11275245 and 11705048, and the ARC Centre of Excellence for Particle Physics at the Tera-scale, under the grant CE110001004.

	\appendix
	\section{Validations of the analyses at LHC~Run~\rom{2}}\label{sec:appendix}
	In this section, we verify the correctness of our implementation of the needed analyses in the package \texttt{CheckMATE}. For the sake of brevity, we only provide the validation of the most sensitive analyses. In Tab.~\ref{val:039} and Tab.~\ref{val:003}, we compare our cut-flows for the analysis in~\cite{Sirunyan:2017lae} and the analysis in~\cite{CMS-PAS-SUS-17-003} with relevant data provided by experimental groups. The results indicate that our simulations are in good agreement with the analysis of the experimental groups.
	
    \begin{table}[htbp]
    \centering
    \begin{tabular}{c|cccc}
    \hline
    Signal region                 & \multicolumn{4}{c}{\texttt{SRA} and \texttt{SRB}}                                        \\
    Process                       & \multicolumn{4}{c}{Production of $\widetilde{\chi}_2^0\widetilde{\chi}_1^\pm$ decay to WZ}   \\
    Point                         & \multicolumn{4}{c}{$m_{\widetilde{\chi}_2^0} = m_{\widetilde{\chi}_1^{\pm}}=200$ GeV; $m_{\widetilde{\chi}_1^0}=100$ GeV}                       \\
    Generated events		& \multicolumn{4}{c}{100,000}\\
    \hline
    \multirow{2}{*}{Selection}    & \multicolumn{2}{c|}{CMS}                 & \multicolumn{2}{c}{\texttt{CheckMATE}} \\
                              & events & \multicolumn{1}{c|}{efficiency} & events       & efficiency     \\ \hline
    3 tight e, $\mu$ or $\tau_{\rm{h}}$  & 482.20 & \multicolumn{1}{c|}{-}          & 482.20       & -              \\
    $4^{\rm{th}}$ lepton veto               & 481.49 & \multicolumn{1}{c|}{99.9\%}     & 481.853      & 99.9\%         \\
    conversions and low-mass veto & 463.71 & \multicolumn{1}{c|}{96.3\%}     & 459.547      & 95.4\%         \\
    $b$-jet veto                    & 456.68 & \multicolumn{1}{c|}{98.5\%}     & 454.896      & 99.0\%         \\
    $E_{\rm T}^{\rm{miss}} > 50$ GeV        & 317.00 & \multicolumn{1}{c|}{69.4\%}     & 290.691      & 63.9\%         \\
    $M_{\rm{T}} > 100$ GeV            & 111.97 & \multicolumn{1}{c|}{35.3\%}     & 105.877      & 36.4\%         \\
    $M_{\ell\ell} > 75 $ GeV           & 103.49 & \multicolumn{1}{c|}{92.4\%}     & 99.8032      & 94.3\%         \\ \hline
    \end{tabular}
    \caption{\label{val:039}Cut-flow validation for signal region categories \texttt{SRA} and \texttt{SRB} in analysis~\cite{Sirunyan:2017lae}. The yields in ``3 tight e, $\mu$ or $\tau_{\rm{h}}$'' of ``\texttt{CheckMATE}'' are normalized to ``3 tight e, $\mu$ or $\tau_{\rm{h}}$'' of ``CMS''. ``efficiency'' is defined as the ratio of the event number passing though the Cut-flow to the event number of the previous one.}
    \end{table}
	
    \begin{table}[htbp]
    \centering
    \resizebox{\textwidth}{!}{
    \begin{tabular}{l|cccccc}
    \hline
    \multicolumn{1}{c|}{\multirow{2}{*}{Process}} & \multicolumn{6}{c}{$pp\to \widetilde{\tau}^+ \widetilde{\tau}^-$, $\widetilde{\tau}^{\pm} \to \tau^{\pm} \widetilde{\chi}_1^0$,}                                                                  \\
    \multicolumn{1}{c|}{}                         & \multicolumn{6}{c}{$\widetilde{\tau}$ is left-handed helicity dominated.}                                                                                      \\
    \multicolumn{1}{c|}{Generated events}                         & \multicolumn{6}{c}{250,000}                                                                                \\ \hline
    \multicolumn{1}{c|}{Point ($m_{\widetilde{\tau}}, m_{\widetilde{\chi}_1^0}$)}                    & \multicolumn{2}{c|}{(100 GeV, 1 GeV)}         & \multicolumn{2}{c|}{(150 GeV, 1 GeV)}         & \multicolumn{2}{c}{(200 GeV, 1 GeV)} \\
    \multicolumn{1}{c|}{Selection}                & CMS & \multicolumn{1}{c|}{\texttt{CheckMATE}} & CMS & \multicolumn{1}{c|}{\texttt{CheckMATE}} & CMS       & \texttt{CheckMATE}      \\ \hline
    Baseline  							& 52.77	  	& \multicolumn{1}{c|}{54.36}      & 24.55 & \multicolumn{1}{c|}{21.08}      & 11.65     & 9.83      \\
    $\Delta_{\phi}(\tau_1, \tau_2)$		& 51.73   	& \multicolumn{1}{c|}{52.41}      & 23.64 & \multicolumn{1}{c|}{19.35}      & 10.60     & 8.76      \\ \hline
    $M_{\rm{T2}} > 90$ GeV       		& 0.10	  	& \multicolumn{1}{c|}{0.40}       & 1.25  & \multicolumn{1}{c|}{1.91}       & 1.67      & 1.77      \\ \hline
    40 GeV $< M_{\rm{T2}} <$ 90 GeV		& 10.64   	& \multicolumn{1}{c|}{14.70}      & 8.99  & \multicolumn{1}{c|}{7.18}       & 3.49      & 3.29      \\
    $E_{\rm T}^{\rm{miss}} > 50$ GeV		& 9.42    	& \multicolumn{1}{c|}{11.97}      & 8.45  & \multicolumn{1}{c|}{6.57}       & 3.28      & 3.03      \\
    300 GeV $< \Sigma M_{\rm{T}} <$ 350 GeV	& 1.06    	& \multicolumn{1}{c|}{1.34}       & 1.53  & \multicolumn{1}{c|}{1.26}       & 0.65      & 0.65      \\
    $\Sigma M_{\rm{T}} >$ 350 GeV			& 1.69    	& \multicolumn{1}{c|}{2.27}       & 2.91  & \multicolumn{1}{c|}{1.93}       & 1.30      & 1.27      \\ \hline
    \end{tabular}}
    \caption{\label{val:003}Cut-flow validation of~\cite{CMS-PAS-SUS-17-003} for different mass points of the left-handed stau sample.}
    \end{table}

\newpage
\bibliography{sample}
\bibliographystyle{CitationStyle}
\end{document}